\documentclass[twocolumn,showpacs,preprintnumbers]{revtex4}
\usepackage{amsfonts}
\usepackage{amsmath}
\usepackage{graphicx}
\usepackage{amssymb}
\usepackage{color,soul}
\usepackage[T1]{fontenc}
\usepackage{ae,aecompl}
\usepackage{natbib}

\begin{document}

\title{Quantum interferometry with a g-factor-tunable spin qubit}
\author{K.~Ono$^{1,2}$\thanks{%
E-mail address:k-ono@riken.jp}, S.~N.~Shevchenko$^{3,4,5}$, T.~Mori$^{6}$,
S.~Moriyama$^{7}$, Franco~Nori$^{3,8}$}
\affiliation{$^1$Advanced device laboratory, RIKEN, Wako-shi, Saitama 351-0198, Japan}
\affiliation{$^{2}$CEMS, RIKEN, Wako-shi, Saitama 351-0198, Japan}
\affiliation{$^{3}$Theoretical Quantum Physics Laboratory, RIKEN Cluster for Pioneering
Research, Wako-shi, Saitama 351-0198, Japan}
\affiliation{$^{4}$B.~Verkin Institute for Low Temperature Physics and Engineering,
Kharkov 61103, Ukraine}
\affiliation{$^{5}$V.~N.~Karazin Kharkov National University, Kharkov 61022, Ukraine}
\affiliation{$^{6}$Nanoelectronics Research Institute, National Institute of Advanced
Industrial Science and Technology (AIST), Tsukuba, Ibaraki 305-8568, Japan}
\affiliation{$^{7}$International Center for Materials Nanoarchitectonics, National
Institute for Materials Science (NIMS), Tsukuba, Ibaraki 305-0044, Japan}
\affiliation{$^{8}$Department of Physics, The University of Michigan, Ann Arbor, MI
48109-1040, USA}
\date{\today }

\begin{abstract}
We study quantum interference effects of a qubit whose energy levels are
continuously modulated. The qubit is formed by an impurity electron spin in
a silicon tunneling field-effect transistor, and it is read out by spin
blockade in a double-dot configuration. The qubit energy levels are
modulated via its gate-voltage-dependent $g$-factors, with either
rectangular, sinusoidal, or ramp radio-frequency waves. The energy-modulated
qubit is probed by the electron spin resonance. Our results demonstrate the
potential of spin qubit interferometry implemented in a silicon device and
operated at a relatively high temperature.%
\end{abstract}

\pacs{73.63.Kv, 73.23.Hk, 76.30.-v}
\maketitle


\textit{Introduction.}---Sensitive measurement techniques are based on the
interference of waves. The most striking illustration is the recent use of
interferometry for the detection of gravitational waves \cite{Abbott16}. If
in place of classical electromagnetic waves one can use the wave functions
of quantum objects, such techniques can be called quantum interferometry.
This was studied not only for conventional small quantum objects \cite{Suda,
Braun18}, but also for large organic molecules \cite{Gerlich11, Hornberger12}
and micrometer-size superconducting circuits \cite{Oliver05, Sillanpaa06,
Shevchenko10}; see also a recent review article [\onlinecite{Degen17}] for
different realizations and applications in quantum sensing. Since it is
difficult to maintain a coherent superposition of charge states, it might be
more beneficial to use instead the spin degree of freedom \cite{Morton11}.
Interestingly, silicon, the second most abundant element in the Earth's
crust and the base of modern electronics, is an ideal environment for spins
in the solid state \cite{Zwanenburg13}. In this work, we will explore how to
use a single-spin silicon-based qubit for quantum interferometry.

Among other characteristics, for quantum engineering it is important to have
qubits which are \textquotedblleft hot, dense, and
coherent\textquotedblright\ \cite{Vandersypen17}. In this context,
\textquotedblleft hot\textquotedblright\ means working in the
technologically less challenging few-Kelvin regime rather than being cooled
down to the milli-Kelvin domain. \textquotedblleft Dense\textquotedblright\
refers to the possibility to achieve high density of quantum dots or donors
in semiconductors. Another benefit of this platform is its compatibility
with the well-developed complementary-metal-oxide-semiconductor (CMOS)
technology. Even more, it has been shown \cite{Maurand16, Gonzalez-Zalba16,
Ono17} that transistors can behave as quantum dots, in which either charge
or spin qubits are realized.

Quantum systems can be modulated by signals of different shapes, such as
sinusoidal and square-wave signals. The latter allows one to rapidly change
a qubit state from one to another, which we can refer to as latching
modulation of qubit states~\cite{Silveri15, Silveri17}. If this is done with
a period longer than the coherence time, then the response has two separate
peaks, situated at the two resonance frequencies corresponding to the two
states. Increasing the modulation frequency, the coherent response is
displayed as an averaged signal, situated at a frequency between the two
resonance frequencies mentioned above, which is known as motional narrowing.
Both motional averaging and narrowing are known in NMR systems and recently
also studied in superconducting systems \cite{Li13, Pan17}. In this way, by
changing the modulation frequency, namely its ratio to the coherence rate,
one can observe the transition between classical (incoherent) and quantum
(coherent) regimes, as in Refs.~\cite{Fink10, Fedorov11, Li13,
Pietikainen17a}.

In this work, we focus on the time-ensemble behavior of a spin-$1/2$ qubit
and study the effect of continuously modulating the qubit energy. In this
way, we explore the motional averaging not only for the symmetric latching
modulation (which was previously demonstrated in superconducting qubits \cite%
{Li13, Silveri15, Silveri17}), but also in\ the asymmetric regime, where
dwelling in one state is longer than in the other state. A square-wave
modulation with variable duty ratio shows weighted motional averaging. At
low modulation frequency, this is visualized, in the frequency dependence,
by two peaks (with weighted height and width); while at high modulation
frequency there is only one averaged peak. We also demonstrate the
sinusoidal energy modulation of the spin qubit and show the Landau-Zener-St%
\"{u}ckelberg-Majorana (LZSM) interference of the spin resonance signal.
This is the first demonstration of LZSM interference where the temperature
is much higher than the photon energy of the sinusoidal modulation
frequency. For realizations of the low-temperature LZSM interference in
quantum-dot systems see Refs.~\cite{Stehlik12, Forster14, Stehlik16,
Korkusinski17, Bogan18, Chatterjee18, Pasek18, Koski18}.

\begin{figure}[t]
\begin{center}
\includegraphics[width=0.48\textwidth, keepaspectratio]{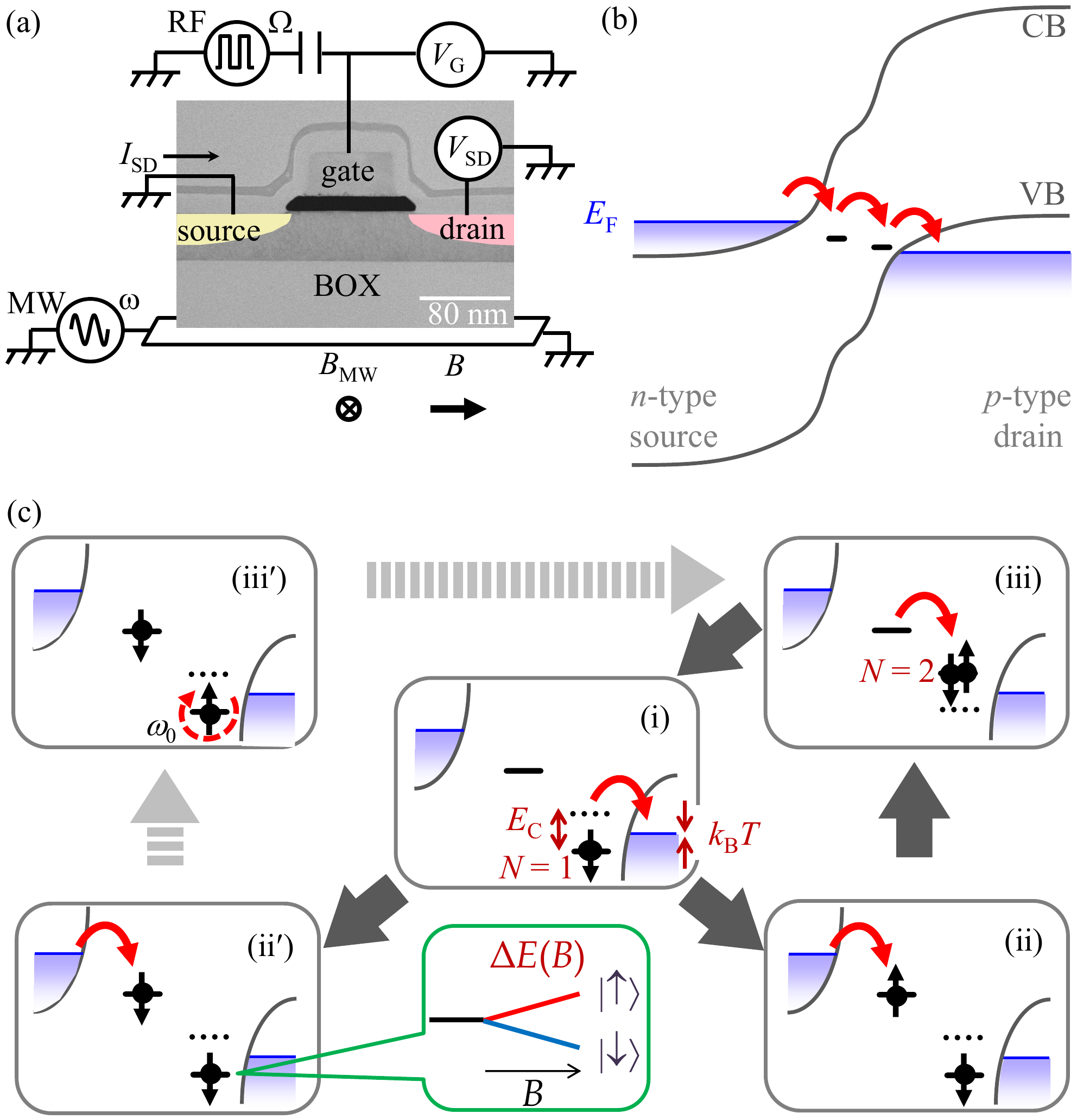}
\end{center}
\caption{\textbf{High-temperature TFET-based single-spin qubit.}
(a)~Schematic of the device and measurement set up. The transistor is
defined on a SOI structure with $n$-type source and $p$-type drain
electrodes. The channel length and width are $80$~nm and $10$~$\protect\mu $%
m, respectively. The source-drain current $I_{SD}$ of the device is measured
for the source-drain voltage $V_{SD}$; and the gate voltage $V_{G}$ at
temperature $T=1.6$~K is achieved with a pumped $^{4}$He cryostat. A
magnetic field $B$ is directed along the source-drain current. A microwave
(MW) signal is applied on the substrate. A gate-voltage modulation in the
MHz regime is applied through a high-pass block capacitor with cutoff
frequency ($<10$~kHz) much smaller than the modulation frequency ($\sim $%
MHz). (b)~Schematic of the potential landscape of the device. (c)~Schematic
of the single-electron tunneling cycle in the spin-blockade regime; see the
SM for details.}
\label{Fig1}
\end{figure}

\textit{Device and measurement.}---We used a spin qubit device based on a
short-channel tunneling field-effect transistor, TFET \cite{Ionescu11}, with
the implanted deep impurity, Fig.~\ref{Fig1}(a) \cite{Mori2014band,
1882-0786-8-3-036503, doi:10.1063/1.4913610, Mori17, Ono18}. The device is
essentially a gate-tunable PIN diode (a diode with an undoped intrinsic
semiconductor region between a $p$-type semiconductor and an $n$-type
semiconductor region). The ion implantations created coupled Al-N impurity
pairs in silicon \cite{weber1980localized, sauer1984nitrogen,
modavis1990aluminum, iizuka2015first}. For an appropriate channel length, a
three-step tunneling from the $n$-type source electrode to the $p$-type
drain electrode occurs via two localized states in the channel, Fig.~\ref%
{Fig1}(b). The PIN structure allows tunneling via the localized states of a
deep impurity and a shallow impurity~\cite{Suppl, Ono18}.

\textit{Spin blockade and ESR.}---The device has two localized states, which
behave as a double quantum dot device, where the current is defined by
single-electron transport \cite{RevModPhys.75.1}. Under an appropriate
source voltage $V_{SD}$ and gate voltage $V_{G}$, the device shows spin
blockade (SB) \cite{ono2002current}. At the electron spin resonance (ESR)
for one of the spins in the double dot, the source-drain current $I_{SD}$
increases due to the lifting of the spin blockade Fig.~\ref{Fig1}(c) \cite%
{koppens2006driven, Maurand16}. Note that the large on-site Coulomb energy
and strong confinement of these impurities allow a spin-qubit operation with
a reasonable coherence time ($T_{2}^{\ast }=0.2-0.3$~$\mu $s) at relatively
high temperatures and low magnetic fields \cite{pla2012single}. Changing the
gate voltage $V_{G}$ within the spin blockade region changes the $g$-factor
by about $1\%$ due to the Stark effect \cite{rahman2009gate}.

We describe our spin qubit device as a two-level system with the pseudo-spin
Hamiltonian $H(t)=B_{z}(t)\sigma _{z}/2+B_{x}(t)\sigma _{x}/2 $. The
longitudinal part is defined by the Zeeman splitting, $B_{z}(t)=g(t)\mu _{%
\mathrm{B}}B$. The time-dependent gate voltage changes the $g$-factor by a
small value and we have $B_{z}/\hbar =\omega _{0}+\delta \cdot s(t)$, where
the amplitude $\delta \ll \omega _{0}$; $\omega _{0}=2\pi f_{0}$ represents
the ESR frequency. In this work, we consider three types of signals \cite%
{Suppl}: a sinusoidal modulation, $s(t)=\cos \Omega t$, a latching
modulation, given by
\begin{equation}
s(t)=\left\{
\begin{array}{c}
2d,\;\;\;\;\;\;\;0<\Omega t/2\pi <1-d, \\
-2\left( 1-d\right) ,\;1-d<\Omega t/2\pi <1,%
\end{array}%
\right.  \label{s(t)}
\end{equation}%
where $d$ is the duty-cycle ratio, and a ramp modulation, given by the
fractional part in $\left\{ \Omega t/2\pi \right\} $. Note that for a
symmetric latching modulation, with $d=1/2$, from Eq.~(\ref{s(t)}), we have $%
s(t)=\mathrm{sgn}\left( \cos \Omega t\right) $. In addition, the transverse
part of the Hamiltonian is defined by the MW voltage applied to the
substrate, $B_{x}/\hbar =2G\cos \omega t$ with amplitude $G$ and circular
frequency $\omega =2\pi f$. The modulation is assumed to be slow, i.e. $%
\Omega \ll \omega $.

\begin{figure}[t]
\begin{center}
\includegraphics[width=0.48 \textwidth, keepaspectratio]{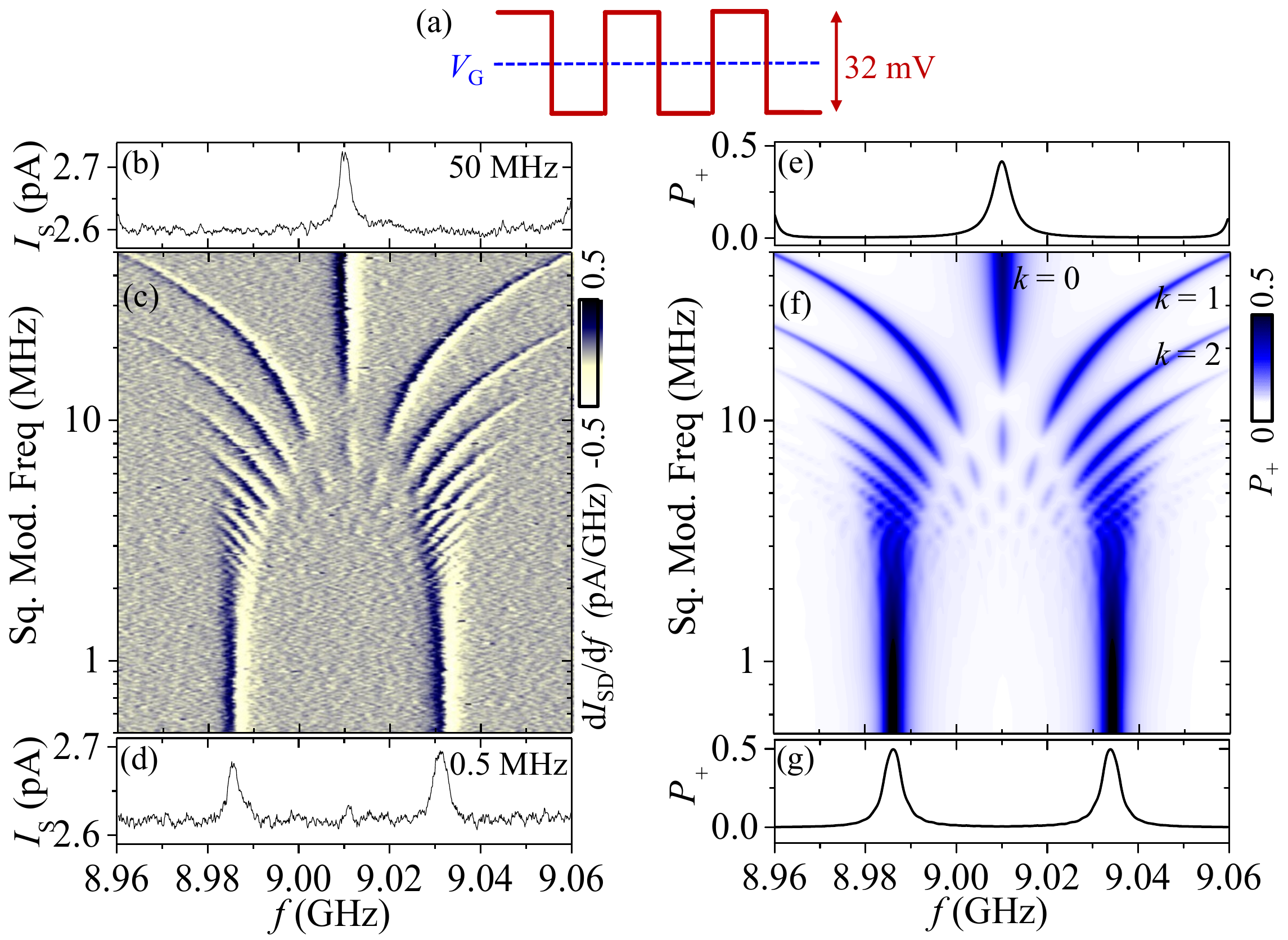}
\end{center}
\caption{\textbf{Square wave modulation of the spin qubit.} (a)~Shape of the
rf signal. (b,c,d) The source-drain current $I_{SD}$ of the device with a
square modulation of its full amplitude $32$~mV added to the gate at $%
V_{G}=-0.36$~V through the high-pass block capacitor. Other conditions are
the same as in Fig.~\protect\ref{Fig1}. In panel~(c) we present the
intensity plot of $dI_{SD}/df$ versus the frequency $f$ and the square-wave
modulation frequency $\Omega $ (log scale from $0.5$ to $50$~MHz), showing
the evolution from the two ESR peaks into the strong main ESR peak and weak
sideband peaks. Note that the distance between the main and the sideband
peaks (seen in the upper area $>10$~MHz) is linear in the modulation
frequency $\Omega $. Panels~(b) and~(d) present the source drain current $%
I_{SD}$ at modulation frequencies of $50$~MHz and $5$~MHz, respectively.
(e,f,g)~Calculation of the upper-state population of the qubit under
square-wave modulation of its energy. For calculations, the following
parameters were used for all the graphs: $G/2\protect\pi =1.1$~MHz, $\Gamma
_{1}/2\protect\pi =0.2$~MHz,\ $\Gamma _{2}/2\protect\pi =1$~MHz, and also
for the right panels (e-g) here: $\protect\delta /2\protect\pi =24$~MHz.}
\label{Fig3}
\end{figure}

\begin{figure}[t]
\begin{center}
\includegraphics[width=0.48 \textwidth, keepaspectratio]{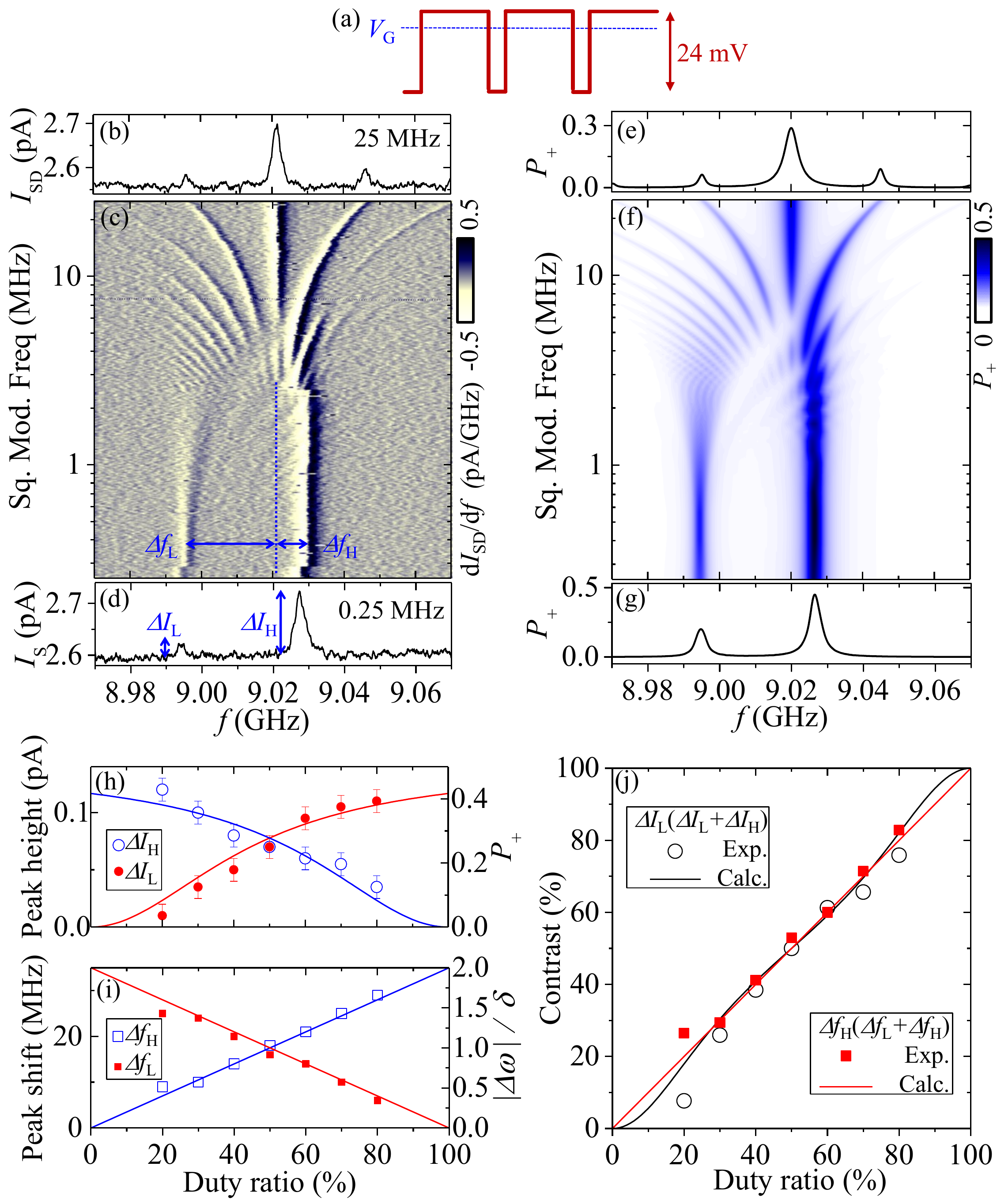}
\end{center}
\caption{ \textbf{Weighted averaging for the spin qubit versus the duty
ratio $d$}\textbf{.} (a)~Shape of a square wave with $d=20\%$. (b-d) Similar
plot of $I_{SD}$ as in Fig.~\protect\ref{Fig3}(b-d) but with a square wave
with a $20\%$ duty ratio, which corresponds to $d=0.2$. The modulation
frequency is changed from $0.25$ to $25$ MHz in a log scale.
(e-g)~Calculation of the qubit upper-state population under square-wave
modulation ($d=0.2$) of its energy. Besides the duty ratio, other parameters
are the same as in Fig.~\protect\ref{Fig3}(e-g). Note that the main weighted
averaged peak appears again at $f=f_{0}$ (here $f_{0}=9.02$ GHz),
independent of the duty ratio. (h)~Duty ratio dependence of the ESR peak
heights at $\Omega /2\protect\pi =0.25$ MHz for lower frequency, $\Delta
I_{L}$, and higher frequency, $\Delta I_{H}$. The $\Delta I_{L}$ and $\Delta
I_{H}$ for $d=0.2$ are indicated in (d). (i)~The duty-ratio dependence among
three ESR peaks, i.e. the motional averaged main peak at $\Omega /2\protect%
\pi =25$~MHz ($9.021$~GHz), to the ESR peaks of lower/higher frequency at $%
\Omega /2\protect\pi =0.25$~MHz. The distances between the peaks, $\Delta
f_{L}$ and $\Delta f_{H}$, are indicated in (e) for $d=0.2$. (j)~Duty-ratio
dependence of the contrasts for the peak heights and distances. }
\label{Fig4}
\end{figure}

\textit{Square-wave modulation.}---In figures \ref{Fig3} and \ref{Fig4} we
present the results of our measurements and calculations for symmetric and
asymmetric square-wave modulation signals. The left panel focuses on the
source-drain current $I_{SD}$, showing the current derivative, $dI_{SD}/df$,
in its main panel. The right panel presents the corresponding theoretical
predictions for the qubit upper-level occupation.

By adding a square-wave MHz modulation signal to the gate [Fig. \ref{Fig3}%
(a)], the gate voltage, i.e. $g$-factor, can be switched between two values,
as described by Eq.~(\ref{s(t)}). Figure~\ref{Fig3}(d) shows $I_{SD}$ at $%
V_{G}$ $=-0.36$~V and square wave of frequency $0.5$~MHz and amplitude $32$%
~mV. Due to the slow measurement with a time constant (that shows how fast
the measured current changes) $\sim 0.3$~s, we observe the two ESR peaks
with two different $g$-factors for $V_{G}$ $=-0.36+0.016$~V and $-0.36-0.016$%
~V, respectively. By increasing the modulation frequency, the ESR peaks show
a characteristic interference pattern and eventually a strong (main) peak
appears with weak sideband peaks [Fig.~\ref{Fig3}(b)]. The strong single
peak at $f=f_{0}=9.01$ GHz [Fig.~\ref{Fig3}(b)] is a result of motional
averaging of the two peaks for the slow modulation [Fig.~\ref{Fig3}(d)]. A
similar pattern was observed for latching modulation of the energy of a
superconducting qubit in Ref.~\cite{Silveri15}. Changing the modulation
amplitude shows a similar behavior with a similar crossover frequency~\cite%
{Suppl}.

In order to describe the system, we solve the Bloch equations with the above
Hamiltonian. We assume that $\omega \gg \Omega $ and after a rotating-wave
approximation%
\begin{equation}
H_{1}=\!\frac{\hbar }{2}\left[ \Delta \omega +\delta \!\cdot \!s(t)\right]
\sigma _{z}+\frac{\hbar G}{2}\sigma _{x},  \label{H_RWA}
\end{equation}%
where $\Delta \omega =\omega _{0}-\omega =2\pi \left( f_{0}-f\right) $.
Details of the calculations are presented in~\cite{Suppl}, cf. Refs.~\cite%
{Shevchenko05, Ivakhnenko18}. As a result, the upper-level occupation
probability is readily obtained from the stationary solution of the Bloch
equations:%
\begin{equation}
P_{+}\!\!\left( \!\Delta \omega ,\frac{\delta }{\Omega }\!\right) \!=\frac{1%
}{2}\sum_{k=-\infty }^{\infty }\frac{G_{k}^{2}(\delta /\Omega )}{%
G_{k}^{2}(\delta /\Omega )\!+\!\frac{\Gamma _{1}}{\Gamma _{2}}\left( \Delta
\omega \!-\!k\Omega \right) ^{2}\!+\!\Gamma _{1}\Gamma _{2}},  \label{Pup}
\end{equation}%
where $G_{k}(x)=G\left\vert \Delta _{k}(x)\right\vert $, which can be
interpreted as the dressed qubit gap, modulated by the function $\Delta
_{k}(x)$. The relaxation and decoherence rates are denoted as $\Gamma
_{1}=T_{1}^{-1}$ and $\Gamma _{2}=T_{2}^{-1}$, respectively. In particular,
for rectangular modulating system with duty-cycle ratio $d$,$\ $we obtain:
\begin{equation}
\left\vert \Delta _{k}(x)\right\vert =\frac{2}{\pi }\frac{x\sin \left[ \pi
\left( 1-d\right) \left( k-2dx\right) \right] }{\left( k+2\left( 1-d\right)
x\right) \left( k-2dx\right) }.  \label{Delta_rect}
\end{equation}

We can interpret the effective Hamiltonian~(\ref{H_RWA}) as follows. The
microwave drive dresses the two-level system resulting in an energy level
difference $\Delta \omega $; when this is matched to the $k$-photon energy
of the rf-signal, the dressed qubit is resonantly excited. Indeed, the
upper-level population in Eq.~(\ref{Pup}) has maxima at $\left\vert \Delta
\omega \right\vert =k\Omega $ \cite{Childress10, Li13}. With Eq.~(3) we
generated the interferograms in the right panels of Figs.~2-4. \cite{Suppl}
In particular, for Fig.~2 we used Eq.~(4) with $d=1/2$.

\textit{Asymmetric modulation.}---Changing the duty ratio $d$ (ratio of the
low $V_{G}$ signal duration to the period; for the previous square wave the
duty ratio was $50$\%), shows both asymmetric modulation and weighted
motional average, as demonstrated in Fig.~\ref{Fig4}. Because the modulation
voltage is added through the block capacitor, the areas of the signal curves
below and above the average gate voltage $V_{G}$ are equal, as shown in Fig.~%
\ref{Fig4}(a). Figure~\ref{Fig4}(d) shows the ESR under slow modulation of
the square-wave signal with a $20$\% duty ratio. The two ESR peak heights
are different, reflecting the duty ratio. For fast modulation, the main peak
appears at the weighted averaged frequency [Fig.~\ref{Fig4}(b)].

We repeat similar measurements with various duty ratios (from $20$ to $80\%$%
). In figure \ref{Fig4}(h) we plot the heights of the two ESR peaks at
lowest modulation frequency. For each duty ratio, we also plot distances
between the above two peak positions and the motional averaged main peak
position at the highest modulation frequency [Fig.~\ref{Fig4}(i)]. Both of
the peak heights and distances reflect the duty ratios. The ratio of the
peak heights and the frequency distances are plotted in Fig.~\ref{Fig4}(j),
and show the motional-averaged main peaks, which indeed appear at the
weighted average frequency. The deviations from linear dependencies in these
plots, especially for the duty ratio of $20\%$, are due to the gate-voltage
dependence of the ESR peak height. More detailed measurements for each duty
ratio are shown in \cite{Suppl}. 

We note here the following interesting features of the weighted motional
averaging. The rectangular-pulse modulation places the qubit in one of the
two allowed positions, and the low-frequency characteristics reflect the
weighted time spent in those two states. For high $\Omega $, the principal
ESR line is situated in-between the two qubit states, the position of which
is independent of the duty ratio. A counter-intuitive aspect is that the
position of this line does not relate to any of the two qubit states, and
thus is referred to as \textquotedblleft motional
averaging\textquotedblright\ \cite{Li13}. Details of calculations are
presented in~\cite{Suppl}. There, it is shown that the frequency shifts are
the following: $\Delta f_{\mathrm{L}}=-2d\delta $ and $\Delta f_{\mathrm{H}%
}=2(1-d)\delta $, while the peak heights are nonlinear functions of $d$.
These formulas are plotted with solid lines in Fig.~\ref{Fig4}(h-j).

\begin{figure}[t]
\begin{center}
\includegraphics[width=0.48 \textwidth, keepaspectratio]{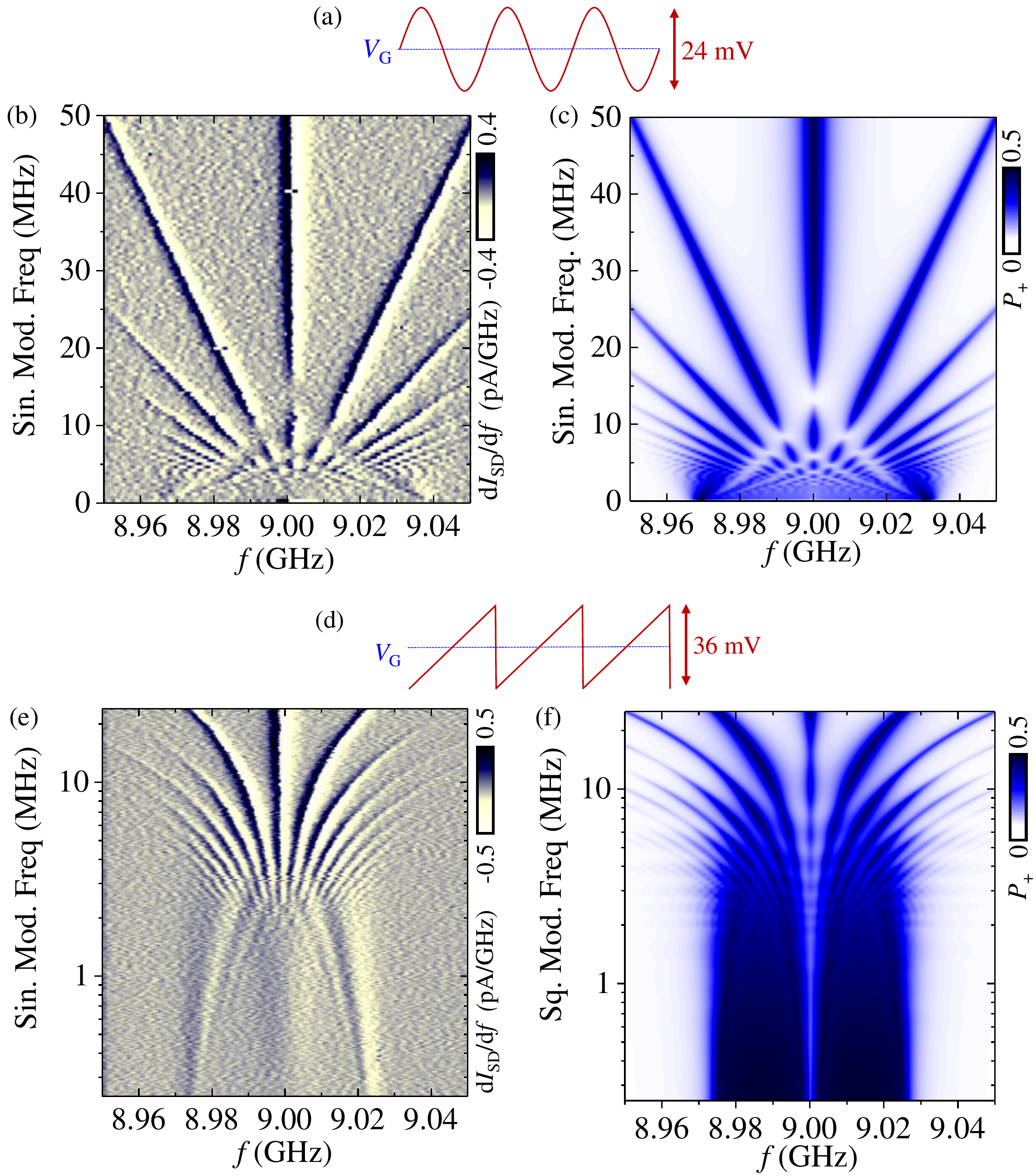}
\end{center}
\caption{\textbf{Sinusoidal and ramp modulation.} (a)~Shape of the rf
sinusoidal signal. (b)~Source-drain current $I_{SD}$ under sinusoidal
modulation with amplitude $24$~mV. Intensity plot of $dI_{SD}/df$ versus the
frequency $f$ and the modulation frequency $\Omega $ (linear scale from $0$
to $80$~MHz). (c)~Calculation of the upper-state population of the qubit
under the sinusoidal modulation of its energy. The parameters are the same
as for Fig.~\protect\ref{Fig3} besides the amplitude, which here is $\protect%
\delta /2\protect\pi =30$~MHz. (d)~ Shape of the ramp wave signal. (e)
Source-drain current $I_{SD}$ under the ramp modulation with amplitude $36$%
~mV. Intensity plot of $dI_{SD}/df$ as a function of the frequency $f$ and
the modulation frequency $\Omega $ (log scale from $0.25$ to $25$~MHz).
(f)~Calculation of the upper-state population of the qubit under ramp
modulation of its energy. The parameters are the same as for Fig.~\protect
\ref{Fig3} besides the amplitude, which here is $\protect\delta /2\protect%
\pi =27$~MHz.}
\label{Fig5}
\end{figure}

\textit{Sinusoidal and ramp modulation.}---Figure \ref{Fig5}(b,c) shows the
effect of the sinusoidal modulation that produces the Landau-Zener-St\"{u}%
ckelberg-Majorana interference patten. The modulation amplitude dependence
with fixed modulation frequency shows the radio-frequency-wave assisted side
bands \cite{Suppl}. In the case of a sinusoidal modulation, the dressed
energy gap is given by the Bessel function of the first kind $\Delta
_{k}(x)=J_{k}(x)$.\cite{Childress10} Then, with Eq.~(\ref{Pup}) we plot Fig.~%
\ref{Fig5}(c). Ramp-wave modulations are shown in Fig.~\ref{Fig5}(e,f). An
inverse ramp waveform gives identical results~\cite{Suppl}. We note that at
low modulation frequency $\Omega $ and low detuning $\Delta \omega $ the
agreement is rather qualitative, which might be due to the rotating-wave
approximation.

\textit{Discussion.}---We have demonstrated that under certain conditions,
an impurity in a field-effect transistor behaves as a single-spin qubit
which displays coherent phenomena, such as Landau-Zener-St\"{u}%
ckelberg-Majorana interference and motional averaging. The spin-qubit device
is based on a short-channel TFET in which, for an appropriate channel
length, a three-step tunneling from the $n$-type source electrode to the $p$%
-type drain electrode occurs via two localized states in the channel. These
localized states (in a deep impurity and a shallow impurity) form a double
quantum dot in which the spin qubit is formed in the spin-blockade regime.
The $g$-factor of the spin can be tuned by the gate voltage, which enables
the fast modulation of the qubit energy. We demonstrated coherent control by
modulating the qubit energy with various continuous waveforms. In
particular, when modulated by asymmetric rectangular pulses with duty ratio $%
d$, we observed interferograms, which we refer to as weighted motional
averaging. At low frequency, this displays $d$-weighted peaks which, at
higher frequency, merge into one peak. To conclude, we summarize the
advantages of the silicon single-spin interferometers: they operate at
relatively high temperature ($1.6~$K), the $g$-factor is controlled by the
gate voltage ($\sim 1\%$, important for selectivity of measurements), the
relaxation times $T_{1,2}$ are large, the fabrication is based on the
well-developed techniques for silicon, such as CMOS, and they can be
manipulated into the ESR and Pauli spin blockade regimes.

\begin{acknowledgments}
We thank K.~Ishibashi for discussions and M.~Cirio and N.~Lambert for
critically reading the manuscript. This work was supported by JSPS KAKENHI
Grant No. 15H04000, 17H01276. F.N. is supported in part by the MURI Center
for Dynamic Magneto-Optics via the Air Force Office of Scientific Research
(AFOSR) (FA9550-14-1-0040), Army Research Office (ARO) (Grant No.
W911NF-18-1-0358), Asian Office of Aerospace Research and Development
(AOARD) (Grant No. FA2386-18-1-4045), Japan Science and Technology Agency
(JST) (Q-LEAP program and CREST Grant No. JPMJCR1676),
Japan Society for the Promotion of Science (JSPS) (JSPS-RFBR Grant No.
17-52-50023 and JSPS-FWO Grant No. VS.059.18N), RIKEN-AIST Challenge
Research Fund, and the John Templeton Foundation.
\end{acknowledgments}

\nocite{apsrev41Control}
\bibliographystyle{apsrev4-1}
\bibliography{references}

\begin{thebibliography}{49}%
\makeatletter
\providecommand \@ifxundefined [1]{%
 \@ifx{#1\undefined}
}%
\providecommand \@ifnum [1]{%
 \ifnum #1\expandafter \@firstoftwo
 \else \expandafter \@secondoftwo
 \fi
}%
\providecommand \@ifx [1]{%
 \ifx #1\expandafter \@firstoftwo
 \else \expandafter \@secondoftwo
 \fi
}%
\providecommand \natexlab [1]{#1}%
\providecommand \enquote  [1]{``#1''}%
\providecommand \bibnamefont  [1]{#1}%
\providecommand \bibfnamefont [1]{#1}%
\providecommand \citenamefont [1]{#1}%
\providecommand \href@noop [0]{\@secondoftwo}%
\providecommand \href [0]{\begingroup \@sanitize@url \@href}%
\providecommand \@href[1]{\@@startlink{#1}\@@href}%
\providecommand \@@href[1]{\endgroup#1\@@endlink}%
\providecommand \@sanitize@url [0]{\catcode `\\12\catcode `\$12\catcode
  `\&12\catcode `\#12\catcode `\^12\catcode `\_12\catcode `\%12\relax}%
\providecommand \@@startlink[1]{}%
\providecommand \@@endlink[0]{}%
\providecommand \url  [0]{\begingroup\@sanitize@url \@url }%
\providecommand \@url [1]{\endgroup\@href {#1}{\urlprefix }}%
\providecommand \urlprefix  [0]{URL }%
\providecommand \Eprint [0]{\href }%
\providecommand \doibase [0]{http://dx.doi.org/}%
\providecommand \selectlanguage [0]{\@gobble}%
\providecommand \bibinfo  [0]{\@secondoftwo}%
\providecommand \bibfield  [0]{\@secondoftwo}%
\providecommand \translation [1]{[#1]}%
\providecommand \BibitemOpen [0]{}%
\providecommand \bibitemStop [0]{}%
\providecommand \bibitemNoStop [0]{.\EOS\space}%
\providecommand \EOS [0]{\spacefactor3000\relax}%
\providecommand \BibitemShut  [1]{\csname bibitem#1\endcsname}%
\let\auto@bib@innerbib\@empty
\bibitem [{\citenamefont {Abbott~\textit{et al}.}(2016)}]{Abbott16}%
  \BibitemOpen
  \bibfield  {author} {\bibinfo {author} {\bibfnamefont {B.~P.}\ \bibnamefont
  {Abbott~\textit{et al}.}} (\bibinfo {collaboration} {LIGO Scientific
  Collaboration and Virgo Collaboration}),\ }\bibfield  {title} {\enquote
  {\bibinfo {title} {Observation of gravitational waves from a binary black
  hole merger},}\ }\href@noop {} {\bibfield  {journal} {\bibinfo  {journal}
  {Phys. Rev. Lett.}\ }\textbf {\bibinfo {volume} {116}},\ \bibinfo {pages}
  {061102} (\bibinfo {year} {2016})}\BibitemShut {NoStop}%
\bibitem [{\citenamefont {Suda}(2006)}]{Suda}%
  \BibitemOpen
  \bibfield  {author} {\bibinfo {author} {\bibfnamefont {M.}~\bibnamefont
  {Suda}},\ }\href@noop {} {\emph {\bibinfo {title} {Quantum Interferometry in
  Phase Space}}}\ (\bibinfo  {publisher} {Springer, Berlin Heidelberg New
  York},\ \bibinfo {year} {2006})\BibitemShut {NoStop}%
\bibitem [{\citenamefont {Braun}\ \emph {et~al.}(2018)\citenamefont {Braun},
  \citenamefont {Adesso}, \citenamefont {Benatti}, \citenamefont {Floreanini},
  \citenamefont {Marzolino}, \citenamefont {Mitchell},\ and\ \citenamefont
  {Pirandola}}]{Braun18}%
  \BibitemOpen
  \bibfield  {author} {\bibinfo {author} {\bibfnamefont {D.}~\bibnamefont
  {Braun}}, \bibinfo {author} {\bibfnamefont {G.}~\bibnamefont {Adesso}},
  \bibinfo {author} {\bibfnamefont {F.}~\bibnamefont {Benatti}}, \bibinfo
  {author} {\bibfnamefont {R.}~\bibnamefont {Floreanini}}, \bibinfo {author}
  {\bibfnamefont {U.}~\bibnamefont {Marzolino}}, \bibinfo {author}
  {\bibfnamefont {M.~W.}\ \bibnamefont {Mitchell}}, \ and\ \bibinfo {author}
  {\bibfnamefont {S.}~\bibnamefont {Pirandola}},\ }\bibfield  {title} {\enquote
  {\bibinfo {title} {Quantum-enhanced measurements without entanglement},}\
  }\href@noop {} {\bibfield  {journal} {\bibinfo  {journal} {Rev. Mod. Phys.}\
  }\textbf {\bibinfo {volume} {90}},\ \bibinfo {pages} {035006} (\bibinfo
  {year} {2018})}\BibitemShut {NoStop}%
\bibitem [{\citenamefont {Gerlich}\ \emph {et~al.}(2011)\citenamefont
  {Gerlich}, \citenamefont {Eibenberger}, \citenamefont {Tomandl},
  \citenamefont {Nimmrichter}, \citenamefont {Hornberger}, \citenamefont
  {Fagan}, \citenamefont {T\"uxen}, \citenamefont {Mayor},\ and\ \citenamefont
  {Arndt}}]{Gerlich11}%
  \BibitemOpen
  \bibfield  {author} {\bibinfo {author} {\bibfnamefont {S.}~\bibnamefont
  {Gerlich}}, \bibinfo {author} {\bibfnamefont {S.}~\bibnamefont
  {Eibenberger}}, \bibinfo {author} {\bibfnamefont {M.}~\bibnamefont
  {Tomandl}}, \bibinfo {author} {\bibfnamefont {S.}~\bibnamefont
  {Nimmrichter}}, \bibinfo {author} {\bibfnamefont {K.}~\bibnamefont
  {Hornberger}}, \bibinfo {author} {\bibfnamefont {P.~J.}\ \bibnamefont
  {Fagan}}, \bibinfo {author} {\bibfnamefont {J.}~\bibnamefont {T\"uxen}},
  \bibinfo {author} {\bibfnamefont {M.}~\bibnamefont {Mayor}}, \ and\ \bibinfo
  {author} {\bibfnamefont {M.}~\bibnamefont {Arndt}},\ }\bibfield  {title}
  {\enquote {\bibinfo {title} {Quantum interference of large organic
  molecules},}\ }\href@noop {} {\bibfield  {journal} {\bibinfo  {journal} {Nat.
  Comm.}\ }\textbf {\bibinfo {volume} {2}},\ \bibinfo {pages} {263} (\bibinfo
  {year} {2011})}\BibitemShut {NoStop}%
\bibitem [{\citenamefont {Hornberger}\ \emph {et~al.}(2012)\citenamefont
  {Hornberger}, \citenamefont {Gerlich}, \citenamefont {Haslinger},
  \citenamefont {Nimmrichter},\ and\ \citenamefont {Arndt}}]{Hornberger12}%
  \BibitemOpen
  \bibfield  {author} {\bibinfo {author} {\bibfnamefont {K.}~\bibnamefont
  {Hornberger}}, \bibinfo {author} {\bibfnamefont {S.}~\bibnamefont {Gerlich}},
  \bibinfo {author} {\bibfnamefont {P.}~\bibnamefont {Haslinger}}, \bibinfo
  {author} {\bibfnamefont {S.}~\bibnamefont {Nimmrichter}}, \ and\ \bibinfo
  {author} {\bibfnamefont {M.}~\bibnamefont {Arndt}},\ }\bibfield  {title}
  {\enquote {\bibinfo {title} {Colloquium: Quantum interference of clusters and
  molecules},}\ }\href@noop {} {\bibfield  {journal} {\bibinfo  {journal} {Rev.
  Mod. Phys.}\ }\textbf {\bibinfo {volume} {84}},\ \bibinfo {pages} {157--173}
  (\bibinfo {year} {2012})}\BibitemShut {NoStop}%
\bibitem [{\citenamefont {Oliver}\ \emph {et~al.}(2005)\citenamefont {Oliver},
  \citenamefont {Yu}, \citenamefont {Lee}, \citenamefont {Berggren},
  \citenamefont {Levitov},\ and\ \citenamefont {Orlando}}]{Oliver05}%
  \BibitemOpen
  \bibfield  {author} {\bibinfo {author} {\bibfnamefont {W.~D.}\ \bibnamefont
  {Oliver}}, \bibinfo {author} {\bibfnamefont {Y.}~\bibnamefont {Yu}}, \bibinfo
  {author} {\bibfnamefont {J.~C.}\ \bibnamefont {Lee}}, \bibinfo {author}
  {\bibfnamefont {K.~K.}\ \bibnamefont {Berggren}}, \bibinfo {author}
  {\bibfnamefont {L.~S.}\ \bibnamefont {Levitov}}, \ and\ \bibinfo {author}
  {\bibfnamefont {T.~P.}\ \bibnamefont {Orlando}},\ }\bibfield  {title}
  {\enquote {\bibinfo {title} {{Mach-Zehnder} interferometry in a strongly
  driven superconducting qubit},}\ }\href@noop {} {\bibfield  {journal}
  {\bibinfo  {journal} {Science}\ }\textbf {\bibinfo {volume} {310}},\ \bibinfo
  {pages} {1653--1657} (\bibinfo {year} {2005})}\BibitemShut {NoStop}%
\bibitem [{\citenamefont {Sillanp\"a\"a}\ \emph {et~al.}(2006)\citenamefont
  {Sillanp\"a\"a}, \citenamefont {Lehtinen}, \citenamefont {Paila},
  \citenamefont {Makhlin},\ and\ \citenamefont {Hakonen}}]{Sillanpaa06}%
  \BibitemOpen
  \bibfield  {author} {\bibinfo {author} {\bibfnamefont {M.}~\bibnamefont
  {Sillanp\"a\"a}}, \bibinfo {author} {\bibfnamefont {T.}~\bibnamefont
  {Lehtinen}}, \bibinfo {author} {\bibfnamefont {A.}~\bibnamefont {Paila}},
  \bibinfo {author} {\bibfnamefont {Y.}~\bibnamefont {Makhlin}}, \ and\
  \bibinfo {author} {\bibfnamefont {P.}~\bibnamefont {Hakonen}},\ }\bibfield
  {title} {\enquote {\bibinfo {title} {Continuous-time monitoring of
  {Landau-Zener} interference in a {C}ooper-pair box},}\ }\href@noop {}
  {\bibfield  {journal} {\bibinfo  {journal} {Phys. Rev. Lett.}\ }\textbf
  {\bibinfo {volume} {96}},\ \bibinfo {pages} {187002} (\bibinfo {year}
  {2006})}\BibitemShut {NoStop}%
\bibitem [{\citenamefont {Shevchenko}\ \emph {et~al.}(2010)\citenamefont
  {Shevchenko}, \citenamefont {Ashhab},\ and\ \citenamefont
  {Nori}}]{Shevchenko10}%
  \BibitemOpen
  \bibfield  {author} {\bibinfo {author} {\bibfnamefont {S.~N.}\ \bibnamefont
  {Shevchenko}}, \bibinfo {author} {\bibfnamefont {S.}~\bibnamefont {Ashhab}},
  \ and\ \bibinfo {author} {\bibfnamefont {F.}~\bibnamefont {Nori}},\
  }\bibfield  {title} {\enquote {\bibinfo {title} {{Landau-Zener-St\"uckelberg}
  interferometry},}\ }\href@noop {} {\bibfield  {journal} {\bibinfo  {journal}
  {Phys. Rep.}\ }\textbf {\bibinfo {volume} {492}},\ \bibinfo {pages} {1--30}
  (\bibinfo {year} {2010})}\BibitemShut {NoStop}%
\bibitem [{\citenamefont {Degen}\ \emph {et~al.}(2017)\citenamefont {Degen},
  \citenamefont {Reinhard},\ and\ \citenamefont {Cappellaro}}]{Degen17}%
  \BibitemOpen
  \bibfield  {author} {\bibinfo {author} {\bibfnamefont {C.~L.}\ \bibnamefont
  {Degen}}, \bibinfo {author} {\bibfnamefont {F.}~\bibnamefont {Reinhard}}, \
  and\ \bibinfo {author} {\bibfnamefont {P.}~\bibnamefont {Cappellaro}},\
  }\bibfield  {title} {\enquote {\bibinfo {title} {Quantum sensing},}\
  }\href@noop {} {\bibfield  {journal} {\bibinfo  {journal} {Rev. Mod. Phys.}\
  }\textbf {\bibinfo {volume} {89}},\ \bibinfo {pages} {035002} (\bibinfo
  {year} {2017})}\BibitemShut {NoStop}%
\bibitem [{\citenamefont {Morton}\ \emph {et~al.}(2011)\citenamefont {Morton},
  \citenamefont {McCamey}, \citenamefont {Eriksson},\ and\ \citenamefont
  {Lyon}}]{Morton11}%
  \BibitemOpen
  \bibfield  {author} {\bibinfo {author} {\bibfnamefont {J.~J.~L.}\
  \bibnamefont {Morton}}, \bibinfo {author} {\bibfnamefont {D.~R.}\
  \bibnamefont {McCamey}}, \bibinfo {author} {\bibfnamefont {M.~A.}\
  \bibnamefont {Eriksson}}, \ and\ \bibinfo {author} {\bibfnamefont {S.~A.}\
  \bibnamefont {Lyon}},\ }\bibfield  {title} {\enquote {\bibinfo {title}
  {Embracing the quantum limit in silicon computing},}\ }\href@noop {}
  {\bibfield  {journal} {\bibinfo  {journal} {Nature}\ }\textbf {\bibinfo
  {volume} {479}},\ \bibinfo {pages} {345--353} (\bibinfo {year}
  {2011})}\BibitemShut {NoStop}%
\bibitem [{\citenamefont {Zwanenburg}\ \emph {et~al.}(2013)\citenamefont
  {Zwanenburg}, \citenamefont {Dzurak}, \citenamefont {Morello}, \citenamefont
  {Simmons}, \citenamefont {Hollenberg}, \citenamefont {Klimeck}, \citenamefont
  {Rogge}, \citenamefont {Coppersmith},\ and\ \citenamefont
  {Eriksson}}]{Zwanenburg13}%
  \BibitemOpen
  \bibfield  {author} {\bibinfo {author} {\bibfnamefont {F.~A.}\ \bibnamefont
  {Zwanenburg}}, \bibinfo {author} {\bibfnamefont {A.~S.}\ \bibnamefont
  {Dzurak}}, \bibinfo {author} {\bibfnamefont {A.}~\bibnamefont {Morello}},
  \bibinfo {author} {\bibfnamefont {M.~Y.}\ \bibnamefont {Simmons}}, \bibinfo
  {author} {\bibfnamefont {L.~C.~L.}\ \bibnamefont {Hollenberg}}, \bibinfo
  {author} {\bibfnamefont {G.}~\bibnamefont {Klimeck}}, \bibinfo {author}
  {\bibfnamefont {S.}~\bibnamefont {Rogge}}, \bibinfo {author} {\bibfnamefont
  {S.~N.}\ \bibnamefont {Coppersmith}}, \ and\ \bibinfo {author} {\bibfnamefont
  {M.~A.}\ \bibnamefont {Eriksson}},\ }\bibfield  {title} {\enquote {\bibinfo
  {title} {Silicon quantum electronics},}\ }\href@noop {} {\bibfield  {journal}
  {\bibinfo  {journal} {Rev. Mod. Phys.}\ }\textbf {\bibinfo {volume} {85}},\
  \bibinfo {pages} {961--1019} (\bibinfo {year} {2013})}\BibitemShut {NoStop}%
\bibitem [{\citenamefont {Vandersypen}\ \emph {et~al.}(2017)\citenamefont
  {Vandersypen}, \citenamefont {Bluhm}, \citenamefont {Clarke}, \citenamefont
  {Dzurak}, \citenamefont {Ishihara}, \citenamefont {Morello}, \citenamefont
  {Reilly}, \citenamefont {Schreiber},\ and\ \citenamefont
  {Veldhorst}}]{Vandersypen17}%
  \BibitemOpen
  \bibfield  {author} {\bibinfo {author} {\bibfnamefont {L.~M.~K.}\
  \bibnamefont {Vandersypen}}, \bibinfo {author} {\bibfnamefont
  {H.}~\bibnamefont {Bluhm}}, \bibinfo {author} {\bibfnamefont {J.~S.}\
  \bibnamefont {Clarke}}, \bibinfo {author} {\bibfnamefont {A.~S.}\
  \bibnamefont {Dzurak}}, \bibinfo {author} {\bibfnamefont {R.}~\bibnamefont
  {Ishihara}}, \bibinfo {author} {\bibfnamefont {A.}~\bibnamefont {Morello}},
  \bibinfo {author} {\bibfnamefont {D.~J.}\ \bibnamefont {Reilly}}, \bibinfo
  {author} {\bibfnamefont {L.~R.}\ \bibnamefont {Schreiber}}, \ and\ \bibinfo
  {author} {\bibfnamefont {M.}~\bibnamefont {Veldhorst}},\ }\bibfield  {title}
  {\enquote {\bibinfo {title} {Interfacing spin qubits in quantum dots and
  donors--hot, dense, and coherent},}\ }\href@noop {} {\bibfield  {journal}
  {\bibinfo  {journal} {npj Quantum Info.}\ }\textbf {\bibinfo {volume} {3}},\
  \bibinfo {pages} {34} (\bibinfo {year} {2017})}\BibitemShut {NoStop}%
\bibitem [{\citenamefont {Maurand}\ \emph {et~al.}(2016)\citenamefont
  {Maurand}, \citenamefont {Jehl}, \citenamefont {Kotekar-Patil}, \citenamefont
  {Corna}, \citenamefont {Bohuslavskyi}, \citenamefont {Lavi\'eville},
  \citenamefont {Hutin}, \citenamefont {Barraud}, \citenamefont {Vinet},
  \citenamefont {Sanquer},\ and\ \citenamefont {De~Franceschi}}]{Maurand16}%
  \BibitemOpen
  \bibfield  {author} {\bibinfo {author} {\bibfnamefont {R.}~\bibnamefont
  {Maurand}}, \bibinfo {author} {\bibfnamefont {X.}~\bibnamefont {Jehl}},
  \bibinfo {author} {\bibfnamefont {D.}~\bibnamefont {Kotekar-Patil}}, \bibinfo
  {author} {\bibfnamefont {A.}~\bibnamefont {Corna}}, \bibinfo {author}
  {\bibfnamefont {H.}~\bibnamefont {Bohuslavskyi}}, \bibinfo {author}
  {\bibfnamefont {R.}~\bibnamefont {Lavi\'eville}}, \bibinfo {author}
  {\bibfnamefont {L.}~\bibnamefont {Hutin}}, \bibinfo {author} {\bibfnamefont
  {S.}~\bibnamefont {Barraud}}, \bibinfo {author} {\bibfnamefont
  {M.}~\bibnamefont {Vinet}}, \bibinfo {author} {\bibfnamefont
  {M.}~\bibnamefont {Sanquer}}, \ and\ \bibinfo {author} {\bibfnamefont
  {S.}~\bibnamefont {De~Franceschi}},\ }\bibfield  {title} {\enquote {\bibinfo
  {title} {A {CMOS} silicon spin qubit},}\ }\href@noop {} {\bibfield  {journal}
  {\bibinfo  {journal} {Nat. Comm.}\ }\textbf {\bibinfo {volume} {7}},\
  \bibinfo {pages} {13575} (\bibinfo {year} {2016})}\BibitemShut {NoStop}%
\bibitem [{\citenamefont {Gonzalez-Zalba}\ \emph {et~al.}(2016)\citenamefont
  {Gonzalez-Zalba}, \citenamefont {Shevchenko}, \citenamefont {Barraud},
  \citenamefont {Johansson}, \citenamefont {Ferguson}, \citenamefont {Nori},\
  and\ \citenamefont {Betz}}]{Gonzalez-Zalba16}%
  \BibitemOpen
  \bibfield  {author} {\bibinfo {author} {\bibfnamefont {M.~F.}\ \bibnamefont
  {Gonzalez-Zalba}}, \bibinfo {author} {\bibfnamefont {S.~N.}\ \bibnamefont
  {Shevchenko}}, \bibinfo {author} {\bibfnamefont {S.}~\bibnamefont {Barraud}},
  \bibinfo {author} {\bibfnamefont {J.~R.}\ \bibnamefont {Johansson}}, \bibinfo
  {author} {\bibfnamefont {A.~J.}\ \bibnamefont {Ferguson}}, \bibinfo {author}
  {\bibfnamefont {F.}~\bibnamefont {Nori}}, \ and\ \bibinfo {author}
  {\bibfnamefont {A.~C.}\ \bibnamefont {Betz}},\ }\bibfield  {title} {\enquote
  {\bibinfo {title} {Gate-sensing coherent charge oscillations in a silicon
  field-effect transistor},}\ }\href@noop {} {\bibfield  {journal} {\bibinfo
  {journal} {Nano Lett.}\ }\textbf {\bibinfo {volume} {16}},\ \bibinfo {pages}
  {1614--1619} (\bibinfo {year} {2016})}\BibitemShut {NoStop}%
\bibitem [{\citenamefont {Ono}\ \emph {et~al.}(2017)\citenamefont {Ono},
  \citenamefont {Giavaras}, \citenamefont {Tanamoto}, \citenamefont {Ohguro},
  \citenamefont {Hu},\ and\ \citenamefont {Nori}}]{Ono17}%
  \BibitemOpen
  \bibfield  {author} {\bibinfo {author} {\bibfnamefont {K.}~\bibnamefont
  {Ono}}, \bibinfo {author} {\bibfnamefont {G.}~\bibnamefont {Giavaras}},
  \bibinfo {author} {\bibfnamefont {T.}~\bibnamefont {Tanamoto}}, \bibinfo
  {author} {\bibfnamefont {T.}~\bibnamefont {Ohguro}}, \bibinfo {author}
  {\bibfnamefont {X.}~\bibnamefont {Hu}}, \ and\ \bibinfo {author}
  {\bibfnamefont {F.}~\bibnamefont {Nori}},\ }\bibfield  {title} {\enquote
  {\bibinfo {title} {Hole spin resonance and spin-orbit coupling in a silicon
  metal-oxide-semiconductor field-effect transistor},}\ }\href@noop {}
  {\bibfield  {journal} {\bibinfo  {journal} {Phys. Rev. Lett.}\ }\textbf
  {\bibinfo {volume} {119}},\ \bibinfo {pages} {156802} (\bibinfo {year}
  {2017})}\BibitemShut {NoStop}%
\bibitem [{\citenamefont {Silveri}\ \emph {et~al.}(2015)\citenamefont
  {Silveri}, \citenamefont {Kumar}, \citenamefont {Tuorila}, \citenamefont
  {Li}, \citenamefont {Veps\"al\"ainen}, \citenamefont {Thuneberg},\ and\
  \citenamefont {Paraoanu}}]{Silveri15}%
  \BibitemOpen
  \bibfield  {author} {\bibinfo {author} {\bibfnamefont {M.~P.}\ \bibnamefont
  {Silveri}}, \bibinfo {author} {\bibfnamefont {K.~S.}\ \bibnamefont {Kumar}},
  \bibinfo {author} {\bibfnamefont {J.}~\bibnamefont {Tuorila}}, \bibinfo
  {author} {\bibfnamefont {J.}~\bibnamefont {Li}}, \bibinfo {author}
  {\bibfnamefont {A.}~\bibnamefont {Veps\"al\"ainen}}, \bibinfo {author}
  {\bibfnamefont {E.~V.}\ \bibnamefont {Thuneberg}}, \ and\ \bibinfo {author}
  {\bibfnamefont {G.~S.}\ \bibnamefont {Paraoanu}},\ }\bibfield  {title}
  {\enquote {\bibinfo {title} {St\"uckelberg interference in a superconducting
  qubit under periodic latching modulation},}\ }\href@noop {} {\bibfield
  {journal} {\bibinfo  {journal} {New J. Phys.}\ }\textbf {\bibinfo {volume}
  {17}},\ \bibinfo {pages} {043058} (\bibinfo {year} {2015})}\BibitemShut
  {NoStop}%
\bibitem [{\citenamefont {Silveri}\ \emph {et~al.}(2017)\citenamefont
  {Silveri}, \citenamefont {Tuorila}, \citenamefont {Thuneberg},\ and\
  \citenamefont {Paraoanu}}]{Silveri17}%
  \BibitemOpen
  \bibfield  {author} {\bibinfo {author} {\bibfnamefont {M.~P.}\ \bibnamefont
  {Silveri}}, \bibinfo {author} {\bibfnamefont {J.~A.}\ \bibnamefont
  {Tuorila}}, \bibinfo {author} {\bibfnamefont {E.~V.}\ \bibnamefont
  {Thuneberg}}, \ and\ \bibinfo {author} {\bibfnamefont {G.~S.}\ \bibnamefont
  {Paraoanu}},\ }\bibfield  {title} {\enquote {\bibinfo {title} {Quantum
  systems under frequency modulation},}\ }\href@noop {} {\bibfield  {journal}
  {\bibinfo  {journal} {Rep. Prog. Phys.}\ }\textbf {\bibinfo {volume} {80}},\
  \bibinfo {pages} {056002} (\bibinfo {year} {2017})}\BibitemShut {NoStop}%
\bibitem [{\citenamefont {Li}\ \emph {et~al.}(2013)\citenamefont {Li},
  \citenamefont {Silveri}, \citenamefont {Kumar}, \citenamefont {Pirkkalainen},
  \citenamefont {Veps\"al\"ainen}, \citenamefont {Chien}, \citenamefont
  {Tuorila}, \citenamefont {Sillanp\"a\"a}, \citenamefont {Hakonen},
  \citenamefont {Thuneberg},\ and\ \citenamefont {Paraoanu}}]{Li13}%
  \BibitemOpen
  \bibfield  {author} {\bibinfo {author} {\bibfnamefont {J.}~\bibnamefont
  {Li}}, \bibinfo {author} {\bibfnamefont {M.~P.}\ \bibnamefont {Silveri}},
  \bibinfo {author} {\bibfnamefont {K.~S.}\ \bibnamefont {Kumar}}, \bibinfo
  {author} {\bibfnamefont {J.-M.}\ \bibnamefont {Pirkkalainen}}, \bibinfo
  {author} {\bibfnamefont {A.}~\bibnamefont {Veps\"al\"ainen}}, \bibinfo
  {author} {\bibfnamefont {W.~C.}\ \bibnamefont {Chien}}, \bibinfo {author}
  {\bibfnamefont {J.}~\bibnamefont {Tuorila}}, \bibinfo {author} {\bibfnamefont
  {M.~A.}\ \bibnamefont {Sillanp\"a\"a}}, \bibinfo {author} {\bibfnamefont
  {P.~J.}\ \bibnamefont {Hakonen}}, \bibinfo {author} {\bibfnamefont {E.~V.}\
  \bibnamefont {Thuneberg}}, \ and\ \bibinfo {author} {\bibfnamefont {G.~S.}\
  \bibnamefont {Paraoanu}},\ }\bibfield  {title} {\enquote {\bibinfo {title}
  {Motional averaging in a superconducting qubit},}\ }\href@noop {} {\bibfield
  {journal} {\bibinfo  {journal} {Nat. Comm.}\ }\textbf {\bibinfo {volume}
  {4}},\ \bibinfo {pages} {1420} (\bibinfo {year} {2013})}\BibitemShut
  {NoStop}%
\bibitem [{\citenamefont {Pan}\ \emph {et~al.}(2017)\citenamefont {Pan},
  \citenamefont {Fan}, \citenamefont {Li}, \citenamefont {Dai}, \citenamefont
  {Wei}, \citenamefont {Lu}, \citenamefont {Cao}, \citenamefont {Kang},
  \citenamefont {Xu}, \citenamefont {Chen}, \citenamefont {Sun},\ and\
  \citenamefont {Wu}}]{Pan17}%
  \BibitemOpen
  \bibfield  {author} {\bibinfo {author} {\bibfnamefont {J.}~\bibnamefont
  {Pan}}, \bibinfo {author} {\bibfnamefont {Y.}~\bibnamefont {Fan}}, \bibinfo
  {author} {\bibfnamefont {Y.}~\bibnamefont {Li}}, \bibinfo {author}
  {\bibfnamefont {X.}~\bibnamefont {Dai}}, \bibinfo {author} {\bibfnamefont
  {X.}~\bibnamefont {Wei}}, \bibinfo {author} {\bibfnamefont {Y.}~\bibnamefont
  {Lu}}, \bibinfo {author} {\bibfnamefont {C.}~\bibnamefont {Cao}}, \bibinfo
  {author} {\bibfnamefont {L.}~\bibnamefont {Kang}}, \bibinfo {author}
  {\bibfnamefont {W.}~\bibnamefont {Xu}}, \bibinfo {author} {\bibfnamefont
  {J.}~\bibnamefont {Chen}}, \bibinfo {author} {\bibfnamefont {G.}~\bibnamefont
  {Sun}}, \ and\ \bibinfo {author} {\bibfnamefont {P.}~\bibnamefont {Wu}},\
  }\bibfield  {title} {\enquote {\bibinfo {title} {Dynamically modulated
  {Autler-Townes} effect in a transmon qubit},}\ }\href@noop {} {\bibfield
  {journal} {\bibinfo  {journal} {Phys. Rev. B}\ }\textbf {\bibinfo {volume}
  {96}},\ \bibinfo {pages} {024502} (\bibinfo {year} {2017})}\BibitemShut
  {NoStop}%
\bibitem [{\citenamefont {Fink}\ \emph {et~al.}(2010)\citenamefont {Fink},
  \citenamefont {Steffen}, \citenamefont {Studer}, \citenamefont {Bishop},
  \citenamefont {Baur}, \citenamefont {Bianchetti}, \citenamefont {Bozyigit},
  \citenamefont {Lang}, \citenamefont {Filipp}, \citenamefont {Leek},\ and\
  \citenamefont {Wallraff}}]{Fink10}%
  \BibitemOpen
  \bibfield  {author} {\bibinfo {author} {\bibfnamefont {J.~M.}\ \bibnamefont
  {Fink}}, \bibinfo {author} {\bibfnamefont {L.}~\bibnamefont {Steffen}},
  \bibinfo {author} {\bibfnamefont {P.}~\bibnamefont {Studer}}, \bibinfo
  {author} {\bibfnamefont {L.~S.}\ \bibnamefont {Bishop}}, \bibinfo {author}
  {\bibfnamefont {M.}~\bibnamefont {Baur}}, \bibinfo {author} {\bibfnamefont
  {R.}~\bibnamefont {Bianchetti}}, \bibinfo {author} {\bibfnamefont
  {D.}~\bibnamefont {Bozyigit}}, \bibinfo {author} {\bibfnamefont
  {C.}~\bibnamefont {Lang}}, \bibinfo {author} {\bibfnamefont {S.}~\bibnamefont
  {Filipp}}, \bibinfo {author} {\bibfnamefont {P.~J.}\ \bibnamefont {Leek}}, \
  and\ \bibinfo {author} {\bibfnamefont {A.}~\bibnamefont {Wallraff}},\
  }\bibfield  {title} {\enquote {\bibinfo {title} {Quantum-to-classical
  transition in cavity quantum electrodynamics},}\ }\href@noop {} {\bibfield
  {journal} {\bibinfo  {journal} {Phys. Rev. Lett.}\ }\textbf {\bibinfo
  {volume} {105}},\ \bibinfo {pages} {163601} (\bibinfo {year}
  {2010})}\BibitemShut {NoStop}%
\bibitem [{\citenamefont {Fedorov}\ \emph {et~al.}(2011)\citenamefont
  {Fedorov}, \citenamefont {Macha}, \citenamefont {Feofanov}, \citenamefont
  {Harmans},\ and\ \citenamefont {Mooij}}]{Fedorov11}%
  \BibitemOpen
  \bibfield  {author} {\bibinfo {author} {\bibfnamefont {A.}~\bibnamefont
  {Fedorov}}, \bibinfo {author} {\bibfnamefont {P.}~\bibnamefont {Macha}},
  \bibinfo {author} {\bibfnamefont {A.~K.}\ \bibnamefont {Feofanov}}, \bibinfo
  {author} {\bibfnamefont {C.~J. P.~M.}\ \bibnamefont {Harmans}}, \ and\
  \bibinfo {author} {\bibfnamefont {J.~E.}\ \bibnamefont {Mooij}},\ }\bibfield
  {title} {\enquote {\bibinfo {title} {Tuned transition from quantum to
  classical for macroscopic quantum states},}\ }\href@noop {} {\bibfield
  {journal} {\bibinfo  {journal} {Phys. Rev. Lett.}\ }\textbf {\bibinfo
  {volume} {106}},\ \bibinfo {pages} {170404} (\bibinfo {year}
  {2011})}\BibitemShut {NoStop}%
\bibitem [{\citenamefont {Pietik\"ainen}\ \emph {et~al.}(2017)\citenamefont
  {Pietik\"ainen}, \citenamefont {Danilin}, \citenamefont {Kumar},
  \citenamefont {Veps\"al\"ainen}, \citenamefont {Golubev}, \citenamefont
  {Tuorila},\ and\ \citenamefont {Paraoanu}}]{Pietikainen17a}%
  \BibitemOpen
  \bibfield  {author} {\bibinfo {author} {\bibfnamefont {I.}~\bibnamefont
  {Pietik\"ainen}}, \bibinfo {author} {\bibfnamefont {S.}~\bibnamefont
  {Danilin}}, \bibinfo {author} {\bibfnamefont {K.~S.}\ \bibnamefont {Kumar}},
  \bibinfo {author} {\bibfnamefont {A.}~\bibnamefont {Veps\"al\"ainen}},
  \bibinfo {author} {\bibfnamefont {D.~S.}\ \bibnamefont {Golubev}}, \bibinfo
  {author} {\bibfnamefont {J.}~\bibnamefont {Tuorila}}, \ and\ \bibinfo
  {author} {\bibfnamefont {G.~S.}\ \bibnamefont {Paraoanu}},\ }\bibfield
  {title} {\enquote {\bibinfo {title} {Observation of the {Bloch-Siegert} shift
  in a driven quantum-to-classical transition},}\ }\href@noop {} {\bibfield
  {journal} {\bibinfo  {journal} {Phys. Rev. B}\ }\textbf {\bibinfo {volume}
  {96}},\ \bibinfo {pages} {020501} (\bibinfo {year} {2017})}\BibitemShut
  {NoStop}%
\bibitem [{\citenamefont {Stehlik}\ \emph {et~al.}(2012)\citenamefont
  {Stehlik}, \citenamefont {Dovzhenko}, \citenamefont {Petta}, \citenamefont
  {Johansson}, \citenamefont {Nori}, \citenamefont {Lu},\ and\ \citenamefont
  {Gossard}}]{Stehlik12}%
  \BibitemOpen
  \bibfield  {author} {\bibinfo {author} {\bibfnamefont {J.}~\bibnamefont
  {Stehlik}}, \bibinfo {author} {\bibfnamefont {Y.}~\bibnamefont {Dovzhenko}},
  \bibinfo {author} {\bibfnamefont {J.~R.}\ \bibnamefont {Petta}}, \bibinfo
  {author} {\bibfnamefont {J.~R.}\ \bibnamefont {Johansson}}, \bibinfo {author}
  {\bibfnamefont {F.}~\bibnamefont {Nori}}, \bibinfo {author} {\bibfnamefont
  {H.}~\bibnamefont {Lu}}, \ and\ \bibinfo {author} {\bibfnamefont {A.~C.}\
  \bibnamefont {Gossard}},\ }\bibfield  {title} {\enquote {\bibinfo {title}
  {{Landau-Zener-St\"uckelberg} interferometry of a single electron charge
  qubit},}\ }\href@noop {} {\bibfield  {journal} {\bibinfo  {journal} {Phys.
  Rev. B}\ }\textbf {\bibinfo {volume} {86}},\ \bibinfo {pages} {121303}
  (\bibinfo {year} {2012})}\BibitemShut {NoStop}%
\bibitem [{\citenamefont {Forster}\ \emph {et~al.}(2014)\citenamefont
  {Forster}, \citenamefont {Petersen}, \citenamefont {Manus}, \citenamefont
  {H\"anggi}, \citenamefont {Schuh}, \citenamefont {Wegscheider}, \citenamefont
  {Kohler},\ and\ \citenamefont {Ludwig}}]{Forster14}%
  \BibitemOpen
  \bibfield  {author} {\bibinfo {author} {\bibfnamefont {F.}~\bibnamefont
  {Forster}}, \bibinfo {author} {\bibfnamefont {G.}~\bibnamefont {Petersen}},
  \bibinfo {author} {\bibfnamefont {S.}~\bibnamefont {Manus}}, \bibinfo
  {author} {\bibfnamefont {P.}~\bibnamefont {H\"anggi}}, \bibinfo {author}
  {\bibfnamefont {D.}~\bibnamefont {Schuh}}, \bibinfo {author} {\bibfnamefont
  {W.}~\bibnamefont {Wegscheider}}, \bibinfo {author} {\bibfnamefont
  {S.}~\bibnamefont {Kohler}}, \ and\ \bibinfo {author} {\bibfnamefont
  {S.}~\bibnamefont {Ludwig}},\ }\bibfield  {title} {\enquote {\bibinfo {title}
  {Characterization of qubit dephasing by {Landau-Zener-St\"uckelberg-Majorana}
  interferometry},}\ }\href@noop {} {\bibfield  {journal} {\bibinfo  {journal}
  {Phys. Rev. Lett.}\ }\textbf {\bibinfo {volume} {112}},\ \bibinfo {pages}
  {116803} (\bibinfo {year} {2014})}\BibitemShut {NoStop}%
\bibitem [{\citenamefont {Stehlik}\ \emph {et~al.}(2016)\citenamefont
  {Stehlik}, \citenamefont {Maialle}, \citenamefont {Degani},\ and\
  \citenamefont {Petta}}]{Stehlik16}%
  \BibitemOpen
  \bibfield  {author} {\bibinfo {author} {\bibfnamefont {J.}~\bibnamefont
  {Stehlik}}, \bibinfo {author} {\bibfnamefont {M.~Z.}\ \bibnamefont
  {Maialle}}, \bibinfo {author} {\bibfnamefont {M.~H.}\ \bibnamefont {Degani}},
  \ and\ \bibinfo {author} {\bibfnamefont {J.~R.}\ \bibnamefont {Petta}},\
  }\bibfield  {title} {\enquote {\bibinfo {title} {Role of multilevel
  {Landau-Zener} interference in extreme harmonic generation},}\ }\href@noop {}
  {\bibfield  {journal} {\bibinfo  {journal} {Phys. Rev. B}\ }\textbf {\bibinfo
  {volume} {94}},\ \bibinfo {pages} {075307} (\bibinfo {year}
  {2016})}\BibitemShut {NoStop}%
\bibitem [{\citenamefont {Korkusinski}\ \emph {et~al.}(2017)\citenamefont
  {Korkusinski}, \citenamefont {Studenikin}, \citenamefont {Aers},
  \citenamefont {Granger}, \citenamefont {Kam},\ and\ \citenamefont
  {Sachrajda}}]{Korkusinski17}%
  \BibitemOpen
  \bibfield  {author} {\bibinfo {author} {\bibfnamefont {M.}~\bibnamefont
  {Korkusinski}}, \bibinfo {author} {\bibfnamefont {S.~A.}\ \bibnamefont
  {Studenikin}}, \bibinfo {author} {\bibfnamefont {G.}~\bibnamefont {Aers}},
  \bibinfo {author} {\bibfnamefont {G.}~\bibnamefont {Granger}}, \bibinfo
  {author} {\bibfnamefont {A.}~\bibnamefont {Kam}}, \ and\ \bibinfo {author}
  {\bibfnamefont {A.~S.}\ \bibnamefont {Sachrajda}},\ }\bibfield  {title}
  {\enquote {\bibinfo {title} {{Landau-Zener-St\"uckelberg} interferometry in
  quantum dots with fast rise times: Evidence for coherent phonon driving},}\
  }\href@noop {} {\bibfield  {journal} {\bibinfo  {journal} {Phys. Rev. Lett.}\
  }\textbf {\bibinfo {volume} {118}},\ \bibinfo {pages} {067701} (\bibinfo
  {year} {2017})}\BibitemShut {NoStop}%
\bibitem [{\citenamefont {Bogan}\ \emph {et~al.}(2018)\citenamefont {Bogan},
  \citenamefont {Studenikin}, \citenamefont {Korkusinski}, \citenamefont
  {Gaudreau}, \citenamefont {Zawadzki}, \citenamefont {Sachrajda},
  \citenamefont {Tracy}, \citenamefont {Reno},\ and\ \citenamefont
  {Hargett}}]{Bogan18}%
  \BibitemOpen
  \bibfield  {author} {\bibinfo {author} {\bibfnamefont {A.}~\bibnamefont
  {Bogan}}, \bibinfo {author} {\bibfnamefont {S.}~\bibnamefont {Studenikin}},
  \bibinfo {author} {\bibfnamefont {M.}~\bibnamefont {Korkusinski}}, \bibinfo
  {author} {\bibfnamefont {L.}~\bibnamefont {Gaudreau}}, \bibinfo {author}
  {\bibfnamefont {P.}~\bibnamefont {Zawadzki}}, \bibinfo {author}
  {\bibfnamefont {A.~S.}\ \bibnamefont {Sachrajda}}, \bibinfo {author}
  {\bibfnamefont {L.}~\bibnamefont {Tracy}}, \bibinfo {author} {\bibfnamefont
  {J.}~\bibnamefont {Reno}}, \ and\ \bibinfo {author} {\bibfnamefont
  {T.}~\bibnamefont {Hargett}},\ }\bibfield  {title} {\enquote {\bibinfo
  {title} {{Landau-Zener-St\"uckelberg-Majorana} interferometry of a single
  hole},}\ }\href@noop {} {\bibfield  {journal} {\bibinfo  {journal} {Phys.
  Rev. Lett.}\ }\textbf {\bibinfo {volume} {120}},\ \bibinfo {pages} {207701}
  (\bibinfo {year} {2018})}\BibitemShut {NoStop}%
\bibitem [{\citenamefont {Chatterjee}\ \emph {et~al.}(2018)\citenamefont
  {Chatterjee}, \citenamefont {Shevchenko}, \citenamefont {Barraud},
  \citenamefont {Otxoa}, \citenamefont {Nori}, \citenamefont {Morton},\ and\
  \citenamefont {Gonzalez-Zalba}}]{Chatterjee18}%
  \BibitemOpen
  \bibfield  {author} {\bibinfo {author} {\bibfnamefont {A.}~\bibnamefont
  {Chatterjee}}, \bibinfo {author} {\bibfnamefont {S.~N.}\ \bibnamefont
  {Shevchenko}}, \bibinfo {author} {\bibfnamefont {S.}~\bibnamefont {Barraud}},
  \bibinfo {author} {\bibfnamefont {R.~M.}\ \bibnamefont {Otxoa}}, \bibinfo
  {author} {\bibfnamefont {F.}~\bibnamefont {Nori}}, \bibinfo {author}
  {\bibfnamefont {J.~J.~L.}\ \bibnamefont {Morton}}, \ and\ \bibinfo {author}
  {\bibfnamefont {M.~F.}\ \bibnamefont {Gonzalez-Zalba}},\ }\bibfield  {title}
  {\enquote {\bibinfo {title} {A silicon-based single-electron interferometer
  coupled to a fermionic sea},}\ }\href@noop {} {\bibfield  {journal} {\bibinfo
   {journal} {Phys. Rev. B}\ }\textbf {\bibinfo {volume} {97}},\ \bibinfo
  {pages} {045405} (\bibinfo {year} {2018})}\BibitemShut {NoStop}%
\bibitem [{\citenamefont {Pasek}\ \emph {et~al.}(2018)\citenamefont {Pasek},
  \citenamefont {Maialle},\ and\ \citenamefont {Degani}}]{Pasek18}%
  \BibitemOpen
  \bibfield  {author} {\bibinfo {author} {\bibfnamefont {W.~J.}\ \bibnamefont
  {Pasek}}, \bibinfo {author} {\bibfnamefont {M.~Z.}\ \bibnamefont {Maialle}},
  \ and\ \bibinfo {author} {\bibfnamefont {M.~H.}\ \bibnamefont {Degani}},\
  }\bibfield  {title} {\enquote {\bibinfo {title} {Application of the
  {Landau-Zener-St\"uckelberg-Majorana} dynamics to the electrically driven
  flip of a hole spin},}\ }\href@noop {} {\bibfield  {journal} {\bibinfo
  {journal} {Phys. Rev. B}\ }\textbf {\bibinfo {volume} {97}},\ \bibinfo
  {pages} {115417} (\bibinfo {year} {2018})}\BibitemShut {NoStop}%
\bibitem [{\citenamefont {Koski}\ \emph {et~al.}(2018)\citenamefont {Koski},
  \citenamefont {Landig}, \citenamefont {Palyi}, \citenamefont {Scarlino},
  \citenamefont {Reichl}, \citenamefont {Wegscheider}, \citenamefont {Burkard},
  \citenamefont {Wallraff}, \citenamefont {Ensslin},\ and\ \citenamefont
  {Ihn}}]{Koski18}%
  \BibitemOpen
  \bibfield  {author} {\bibinfo {author} {\bibfnamefont {J.~V.}\ \bibnamefont
  {Koski}}, \bibinfo {author} {\bibfnamefont {A.~J.}\ \bibnamefont {Landig}},
  \bibinfo {author} {\bibfnamefont {A.}~\bibnamefont {Palyi}}, \bibinfo
  {author} {\bibfnamefont {P.}~\bibnamefont {Scarlino}}, \bibinfo {author}
  {\bibfnamefont {C.}~\bibnamefont {Reichl}}, \bibinfo {author} {\bibfnamefont
  {W.}~\bibnamefont {Wegscheider}}, \bibinfo {author} {\bibfnamefont
  {G.}~\bibnamefont {Burkard}}, \bibinfo {author} {\bibfnamefont
  {A.}~\bibnamefont {Wallraff}}, \bibinfo {author} {\bibfnamefont
  {K.}~\bibnamefont {Ensslin}}, \ and\ \bibinfo {author} {\bibfnamefont
  {T.}~\bibnamefont {Ihn}},\ }\bibfield  {title} {\enquote {\bibinfo {title}
  {Floquet spectroscopy of a strongly driven quantum dot charge qubit with a
  microwave resonator},}\ }\href@noop {} {\bibfield  {journal} {\bibinfo
  {journal} {Phys. Rev. Lett.}\ }\textbf {\bibinfo {volume} {121}},\ \bibinfo
  {pages} {043603} (\bibinfo {year} {2018})}\BibitemShut {NoStop}%
\bibitem [{\citenamefont {Ionescu}\ and\ \citenamefont
  {Riel}(2011)}]{Ionescu11}%
  \BibitemOpen
  \bibfield  {author} {\bibinfo {author} {\bibfnamefont {A.~M.}\ \bibnamefont
  {Ionescu}}\ and\ \bibinfo {author} {\bibfnamefont {H.}~\bibnamefont {Riel}},\
  }\bibfield  {title} {\enquote {\bibinfo {title} {Tunnel field-effect
  transistors as energy-efficient electronic switches},}\ }\href@noop {}
  {\bibfield  {journal} {\bibinfo  {journal} {Nature}\ }\textbf {\bibinfo
  {volume} {479}},\ \bibinfo {pages} {329--337} (\bibinfo {year}
  {2011})}\BibitemShut {NoStop}%
\bibitem [{\citenamefont {Mori}\ \emph {et~al.}(2014)\citenamefont {Mori},
  \citenamefont {Morita}, \citenamefont {Miyata}, \citenamefont {Migita},
  \citenamefont {Fukuda}, \citenamefont {Masahara}, \citenamefont {Yasuda},\
  and\ \citenamefont {Ota}}]{Mori2014band}%
  \BibitemOpen
  \bibfield  {author} {\bibinfo {author} {\bibfnamefont {T.}~\bibnamefont
  {Mori}}, \bibinfo {author} {\bibfnamefont {Y.}~\bibnamefont {Morita}},
  \bibinfo {author} {\bibfnamefont {N.}~\bibnamefont {Miyata}}, \bibinfo
  {author} {\bibfnamefont {S.}~\bibnamefont {Migita}}, \bibinfo {author}
  {\bibfnamefont {K.}~\bibnamefont {Fukuda}}, \bibinfo {author} {\bibfnamefont
  {M.}~\bibnamefont {Masahara}}, \bibinfo {author} {\bibfnamefont
  {T.}~\bibnamefont {Yasuda}}, \ and\ \bibinfo {author} {\bibfnamefont
  {H.}~\bibnamefont {Ota}},\ }\bibfield  {title} {\enquote {\bibinfo {title}
  {Band-to-band tunneling current enhancement utilizing isoelectronic trap and
  its application to {TFETs}},}\ }in\ \href@noop {} {\emph {\bibinfo
  {booktitle} {VLSI Technology (VLSI-Technology): Digest of Technical Papers,
  2014 Symposium on}}}\ (\bibinfo {organization} {IEEE},\ \bibinfo {year}
  {2014})\ pp.\ \bibinfo {pages} {1--2}\BibitemShut {NoStop}%
\bibitem [{\citenamefont {Mori}\ \emph
  {et~al.}(2015{\natexlab{a}})\citenamefont {Mori}, \citenamefont
  {Mizubayashi}, \citenamefont {Morita}, \citenamefont {Migita}, \citenamefont
  {Fukuda}, \citenamefont {Miyata}, \citenamefont {Yasuda}, \citenamefont
  {Masahara},\ and\ \citenamefont {Ota}}]{1882-0786-8-3-036503}%
  \BibitemOpen
  \bibfield  {author} {\bibinfo {author} {\bibfnamefont {T.}~\bibnamefont
  {Mori}}, \bibinfo {author} {\bibfnamefont {W.}~\bibnamefont {Mizubayashi}},
  \bibinfo {author} {\bibfnamefont {Y.}~\bibnamefont {Morita}}, \bibinfo
  {author} {\bibfnamefont {S.}~\bibnamefont {Migita}}, \bibinfo {author}
  {\bibfnamefont {K.}~\bibnamefont {Fukuda}}, \bibinfo {author} {\bibfnamefont
  {N.}~\bibnamefont {Miyata}}, \bibinfo {author} {\bibfnamefont
  {T.}~\bibnamefont {Yasuda}}, \bibinfo {author} {\bibfnamefont
  {M.}~\bibnamefont {Masahara}}, \ and\ \bibinfo {author} {\bibfnamefont
  {H.}~\bibnamefont {Ota}},\ }\bibfield  {title} {\enquote {\bibinfo {title}
  {Effect of hot implantation on {ON}-current enhancement utilizing
  isoelectronic trap in {Si}-based tunnel field-effect transistors},}\
  }\href@noop {} {\bibfield  {journal} {\bibinfo  {journal} {Appl. Phys.
  Expr.}\ }\textbf {\bibinfo {volume} {8}},\ \bibinfo {pages} {036503}
  (\bibinfo {year} {2015}{\natexlab{a}})}\BibitemShut {NoStop}%
\bibitem [{\citenamefont {Mori}\ \emph
  {et~al.}(2015{\natexlab{b}})\citenamefont {Mori}, \citenamefont {Morita},
  \citenamefont {Miyata}, \citenamefont {Migita}, \citenamefont {Fukuda},
  \citenamefont {Mizubayashi}, \citenamefont {Masahara}, \citenamefont
  {Yasuda},\ and\ \citenamefont {Ota}}]{doi:10.1063/1.4913610}%
  \BibitemOpen
  \bibfield  {author} {\bibinfo {author} {\bibfnamefont {T.}~\bibnamefont
  {Mori}}, \bibinfo {author} {\bibfnamefont {Y.}~\bibnamefont {Morita}},
  \bibinfo {author} {\bibfnamefont {N.}~\bibnamefont {Miyata}}, \bibinfo
  {author} {\bibfnamefont {S.}~\bibnamefont {Migita}}, \bibinfo {author}
  {\bibfnamefont {K.}~\bibnamefont {Fukuda}}, \bibinfo {author} {\bibfnamefont
  {W.}~\bibnamefont {Mizubayashi}}, \bibinfo {author} {\bibfnamefont
  {M.}~\bibnamefont {Masahara}}, \bibinfo {author} {\bibfnamefont
  {T.}~\bibnamefont {Yasuda}}, \ and\ \bibinfo {author} {\bibfnamefont
  {H.}~\bibnamefont {Ota}},\ }\bibfield  {title} {\enquote {\bibinfo {title}
  {Study of tunneling transport in {Si}-based tunnel field-effect transistors
  with {ON} current enhancement utilizing isoelectronic trap},}\ }\href@noop {}
  {\bibfield  {journal} {\bibinfo  {journal} {Appl. Phys. Lett.}\ }\textbf
  {\bibinfo {volume} {106}},\ \bibinfo {pages} {083501} (\bibinfo {year}
  {2015}{\natexlab{b}})}\BibitemShut {NoStop}%
\bibitem [{\citenamefont {Mori}\ \emph {et~al.}(2017)\citenamefont {Mori},
  \citenamefont {Iizuka},\ and\ \citenamefont {Nakayama}}]{Mori17}%
  \BibitemOpen
  \bibfield  {author} {\bibinfo {author} {\bibfnamefont {T.}~\bibnamefont
  {Mori}}, \bibinfo {author} {\bibfnamefont {S.}~\bibnamefont {Iizuka}}, \ and\
  \bibinfo {author} {\bibfnamefont {T.}~\bibnamefont {Nakayama}},\ }\bibfield
  {title} {\enquote {\bibinfo {title} {Material engineering for silicon tunnel
  field-effect transistors: isoelectronic trap technology},}\ }\href@noop {}
  {\bibfield  {journal} {\bibinfo  {journal} {MRS Communications}\ }\textbf
  {\bibinfo {volume} {7}},\ \bibinfo {pages} {541--550} (\bibinfo {year}
  {2017})}\BibitemShut {NoStop}%
\bibitem [{\citenamefont {Ono}\ \emph {et~al.}(2019)\citenamefont {Ono},
  \citenamefont {Mori},\ and\ \citenamefont {Moriyama}}]{Ono18}%
  \BibitemOpen
  \bibfield  {author} {\bibinfo {author} {\bibfnamefont {K.}~\bibnamefont
  {Ono}}, \bibinfo {author} {\bibfnamefont {T.}~\bibnamefont {Mori}}, \ and\
  \bibinfo {author} {\bibfnamefont {S.}~\bibnamefont {Moriyama}},\ }\bibfield
  {title} {\enquote {\bibinfo {title} {High-temperature operation of a silicon
  qubit},}\ }\href@noop {} {\bibfield  {journal} {\bibinfo  {journal} {Sci.
  Rep.}\ }\textbf {\bibinfo {volume} {9}},\ \bibinfo {pages} {469} (\bibinfo
  {year} {2019})}\BibitemShut {NoStop}%
\bibitem [{\citenamefont {Weber}\ \emph {et~al.}(1980)\citenamefont {Weber},
  \citenamefont {Schmid},\ and\ \citenamefont {Sauer}}]{weber1980localized}%
  \BibitemOpen
  \bibfield  {author} {\bibinfo {author} {\bibfnamefont {J.}~\bibnamefont
  {Weber}}, \bibinfo {author} {\bibfnamefont {W.}~\bibnamefont {Schmid}}, \
  and\ \bibinfo {author} {\bibfnamefont {R.}~\bibnamefont {Sauer}},\ }\bibfield
   {title} {\enquote {\bibinfo {title} {Localized exciton bound to an
  isoelectronic trap in silicon},}\ }\href@noop {} {\bibfield  {journal}
  {\bibinfo  {journal} {Phys. Rev. B}\ }\textbf {\bibinfo {volume} {21}},\
  \bibinfo {pages} {2401} (\bibinfo {year} {1980})}\BibitemShut {NoStop}%
\bibitem [{\citenamefont {Sauer}\ \emph {et~al.}(1984)\citenamefont {Sauer},
  \citenamefont {Weber},\ and\ \citenamefont {Zulehner}}]{sauer1984nitrogen}%
  \BibitemOpen
  \bibfield  {author} {\bibinfo {author} {\bibfnamefont {R.}~\bibnamefont
  {Sauer}}, \bibinfo {author} {\bibfnamefont {J.}~\bibnamefont {Weber}}, \ and\
  \bibinfo {author} {\bibfnamefont {W.}~\bibnamefont {Zulehner}},\ }\bibfield
  {title} {\enquote {\bibinfo {title} {Nitrogen in silicon: Towards the
  identification of the {1.1223-eV (A, B, C)} photoluminescence lines},}\
  }\href@noop {} {\bibfield  {journal} {\bibinfo  {journal} {Appl. Phys.
  Lett.}\ }\textbf {\bibinfo {volume} {44}},\ \bibinfo {pages} {440--442}
  (\bibinfo {year} {1984})}\BibitemShut {NoStop}%
\bibitem [{\citenamefont {Modavis}\ and\ \citenamefont
  {Hall}(1990)}]{modavis1990aluminum}%
  \BibitemOpen
  \bibfield  {author} {\bibinfo {author} {\bibfnamefont {R.~A.}\ \bibnamefont
  {Modavis}}\ and\ \bibinfo {author} {\bibfnamefont {D.~G.}\ \bibnamefont
  {Hall}},\ }\bibfield  {title} {\enquote {\bibinfo {title} {Aluminum-nitrogen
  isoelectronic trap in silicon},}\ }\href@noop {} {\bibfield  {journal}
  {\bibinfo  {journal} {J. Appl. Phys.}\ }\textbf {\bibinfo {volume} {67}},\
  \bibinfo {pages} {545--547} (\bibinfo {year} {1990})}\BibitemShut {NoStop}%
\bibitem [{\citenamefont {Iizuka}\ and\ \citenamefont
  {Nakayama}(2015)}]{iizuka2015first}%
  \BibitemOpen
  \bibfield  {author} {\bibinfo {author} {\bibfnamefont {S.}~\bibnamefont
  {Iizuka}}\ and\ \bibinfo {author} {\bibfnamefont {T.}~\bibnamefont
  {Nakayama}},\ }\bibfield  {title} {\enquote {\bibinfo {title}
  {First-principles calculation of electronic properties of isoelectronic
  impurity complexes in {Si}},}\ }\href@noop {} {\bibfield  {journal} {\bibinfo
   {journal} {Appl. Phys. Expr.}\ }\textbf {\bibinfo {volume} {8}},\ \bibinfo
  {pages} {081301} (\bibinfo {year} {2015})}\BibitemShut {NoStop}%
\bibitem [{Sup()}]{Suppl}%
  \BibitemOpen
  \href@noop {} {\bibinfo  {journal} {See Supplemental Material at [URL will be
  inserted by publisher]}\ }\BibitemShut {NoStop}%
\bibitem [{\citenamefont {van~der Wiel}\ \emph {et~al.}(2002)\citenamefont
  {van~der Wiel}, \citenamefont {De~Franceschi}, \citenamefont {Elzerman},
  \citenamefont {Fujisawa}, \citenamefont {Tarucha},\ and\ \citenamefont
  {Kouwenhoven}}]{RevModPhys.75.1}%
  \BibitemOpen
\bibfield  {journal} {  }\bibfield  {author} {\bibinfo {author} {\bibfnamefont
  {W.~G.}\ \bibnamefont {van~der Wiel}}, \bibinfo {author} {\bibfnamefont
  {S.}~\bibnamefont {De~Franceschi}}, \bibinfo {author} {\bibfnamefont {J.~M.}\
  \bibnamefont {Elzerman}}, \bibinfo {author} {\bibfnamefont {T.}~\bibnamefont
  {Fujisawa}}, \bibinfo {author} {\bibfnamefont {S.}~\bibnamefont {Tarucha}}, \
  and\ \bibinfo {author} {\bibfnamefont {L.~P.}\ \bibnamefont {Kouwenhoven}},\
  }\bibfield  {title} {\enquote {\bibinfo {title} {Electron transport through
  double quantum dots},}\ }\href@noop {} {\bibfield  {journal} {\bibinfo
  {journal} {Rev. Mod. Phys.}\ }\textbf {\bibinfo {volume} {75}},\ \bibinfo
  {pages} {1--22} (\bibinfo {year} {2002})}\BibitemShut {NoStop}%
\bibitem [{\citenamefont {Ono}\ \emph {et~al.}(2002)\citenamefont {Ono},
  \citenamefont {Austing}, \citenamefont {Tokura},\ and\ \citenamefont
  {Tarucha}}]{ono2002current}%
  \BibitemOpen
  \bibfield  {author} {\bibinfo {author} {\bibfnamefont {K.}~\bibnamefont
  {Ono}}, \bibinfo {author} {\bibfnamefont {D.}~\bibnamefont {Austing}},
  \bibinfo {author} {\bibfnamefont {Y.}~\bibnamefont {Tokura}}, \ and\ \bibinfo
  {author} {\bibfnamefont {S.}~\bibnamefont {Tarucha}},\ }\bibfield  {title}
  {\enquote {\bibinfo {title} {Current rectification by {P}auli exclusion in a
  weakly coupled double quantum dot system},}\ }\href@noop {} {\bibfield
  {journal} {\bibinfo  {journal} {Science}\ }\textbf {\bibinfo {volume}
  {297}},\ \bibinfo {pages} {1313--1317} (\bibinfo {year} {2002})}\BibitemShut
  {NoStop}%
\bibitem [{\citenamefont {Koppens}\ \emph {et~al.}(2006)\citenamefont
  {Koppens}, \citenamefont {Buizert}, \citenamefont {Tielrooij}, \citenamefont
  {Vink}, \citenamefont {Nowack}, \citenamefont {Meunier}, \citenamefont
  {Kouwenhoven},\ and\ \citenamefont {Vandersypen}}]{koppens2006driven}%
  \BibitemOpen
  \bibfield  {author} {\bibinfo {author} {\bibfnamefont {F.~H.~L.}\
  \bibnamefont {Koppens}}, \bibinfo {author} {\bibfnamefont {C.}~\bibnamefont
  {Buizert}}, \bibinfo {author} {\bibfnamefont {K.-J.}\ \bibnamefont
  {Tielrooij}}, \bibinfo {author} {\bibfnamefont {I.~T.}\ \bibnamefont {Vink}},
  \bibinfo {author} {\bibfnamefont {K.~C.}\ \bibnamefont {Nowack}}, \bibinfo
  {author} {\bibfnamefont {T.}~\bibnamefont {Meunier}}, \bibinfo {author}
  {\bibfnamefont {L.~P.}\ \bibnamefont {Kouwenhoven}}, \ and\ \bibinfo {author}
  {\bibfnamefont {L.~M.~K.}\ \bibnamefont {Vandersypen}},\ }\bibfield  {title}
  {\enquote {\bibinfo {title} {Driven coherent oscillations of a single
  electron spin in a quantum dot},}\ }\href@noop {} {\bibfield  {journal}
  {\bibinfo  {journal} {Nature}\ }\textbf {\bibinfo {volume} {442}},\ \bibinfo
  {pages} {766--771} (\bibinfo {year} {2006})}\BibitemShut {NoStop}%
\bibitem [{\citenamefont {Pla}\ \emph {et~al.}(2012)\citenamefont {Pla},
  \citenamefont {Tan}, \citenamefont {Dehollain}, \citenamefont {Lim},
  \citenamefont {Morton}, \citenamefont {Jamieson}, \citenamefont {Dzurak},\
  and\ \citenamefont {Morello}}]{pla2012single}%
  \BibitemOpen
  \bibfield  {author} {\bibinfo {author} {\bibfnamefont {J.~J.}\ \bibnamefont
  {Pla}}, \bibinfo {author} {\bibfnamefont {K.~Y.}\ \bibnamefont {Tan}},
  \bibinfo {author} {\bibfnamefont {J.~P.}\ \bibnamefont {Dehollain}}, \bibinfo
  {author} {\bibfnamefont {W.~H.}\ \bibnamefont {Lim}}, \bibinfo {author}
  {\bibfnamefont {J.~J.~L.}\ \bibnamefont {Morton}}, \bibinfo {author}
  {\bibfnamefont {D.~N.}\ \bibnamefont {Jamieson}}, \bibinfo {author}
  {\bibfnamefont {A.~S.}\ \bibnamefont {Dzurak}}, \ and\ \bibinfo {author}
  {\bibfnamefont {A.}~\bibnamefont {Morello}},\ }\bibfield  {title} {\enquote
  {\bibinfo {title} {A single-atom electron spin qubit in silicon},}\
  }\href@noop {} {\bibfield  {journal} {\bibinfo  {journal} {Nature}\ }\textbf
  {\bibinfo {volume} {489}},\ \bibinfo {pages} {541--545} (\bibinfo {year}
  {2012})}\BibitemShut {NoStop}%
\bibitem [{\citenamefont {Rahman}\ \emph {et~al.}(2009)\citenamefont {Rahman},
  \citenamefont {Park}, \citenamefont {Boykin}, \citenamefont {Klimeck},
  \citenamefont {Rogge},\ and\ \citenamefont {Hollenberg}}]{rahman2009gate}%
  \BibitemOpen
  \bibfield  {author} {\bibinfo {author} {\bibfnamefont {R.}~\bibnamefont
  {Rahman}}, \bibinfo {author} {\bibfnamefont {S.~H.}\ \bibnamefont {Park}},
  \bibinfo {author} {\bibfnamefont {T.~B.}\ \bibnamefont {Boykin}}, \bibinfo
  {author} {\bibfnamefont {G.}~\bibnamefont {Klimeck}}, \bibinfo {author}
  {\bibfnamefont {S.}~\bibnamefont {Rogge}}, \ and\ \bibinfo {author}
  {\bibfnamefont {L.~C.~L.}\ \bibnamefont {Hollenberg}},\ }\bibfield  {title}
  {\enquote {\bibinfo {title} {Gate-induced g-factor control and dimensional
  transition for donors in multivalley semiconductors},}\ }\href@noop {}
  {\bibfield  {journal} {\bibinfo  {journal} {Phys. Rev. B}\ }\textbf {\bibinfo
  {volume} {80}},\ \bibinfo {pages} {155301} (\bibinfo {year}
  {2009})}\BibitemShut {NoStop}%
\bibitem [{\citenamefont {Shevchenko}\ \emph {et~al.}(2005)\citenamefont
  {Shevchenko}, \citenamefont {Kiyko}, \citenamefont {Omelyanchouk},\ and\
  \citenamefont {Krech}}]{Shevchenko05}%
  \BibitemOpen
  \bibfield  {author} {\bibinfo {author} {\bibfnamefont {S.~N.}\ \bibnamefont
  {Shevchenko}}, \bibinfo {author} {\bibfnamefont {A.~S.}\ \bibnamefont
  {Kiyko}}, \bibinfo {author} {\bibfnamefont {A.~N.}\ \bibnamefont
  {Omelyanchouk}}, \ and\ \bibinfo {author} {\bibfnamefont {W.}~\bibnamefont
  {Krech}},\ }\bibfield  {title} {\enquote {\bibinfo {title} {Dynamic behavior
  of {Josephson-junction qubits: crossover between Rabi oscillations and
  Landau-Zener} transitions},}\ }\href@noop {} {\bibfield  {journal} {\bibinfo
  {journal} {Low Temp. Phys.}\ }\textbf {\bibinfo {volume} {31}},\ \bibinfo
  {pages} {569--576} (\bibinfo {year} {2005})}\BibitemShut {NoStop}%
\bibitem [{\citenamefont {Ivakhnenko}\ \emph {et~al.}(2018)\citenamefont
  {Ivakhnenko}, \citenamefont {Shevchenko},\ and\ \citenamefont
  {Nori}}]{Ivakhnenko18}%
  \BibitemOpen
  \bibfield  {author} {\bibinfo {author} {\bibfnamefont {O.~V.}\ \bibnamefont
  {Ivakhnenko}}, \bibinfo {author} {\bibfnamefont {S.~N.}\ \bibnamefont
  {Shevchenko}}, \ and\ \bibinfo {author} {\bibfnamefont {F.}~\bibnamefont
  {Nori}},\ }\bibfield  {title} {\enquote {\bibinfo {title} {Simulating quantum
  dynamical phenomena using classical oscillators:
  {Landau-Zener-St\"uckelberg-Majorana} interferometry, latching modulation,
  and motional averaging},}\ }\href@noop {} {\bibfield  {journal} {\bibinfo
  {journal} {Sci. Rep.}\ }\textbf {\bibinfo {volume} {8}},\ \bibinfo {pages}
  {12218} (\bibinfo {year} {2018})}\BibitemShut {NoStop}%
\bibitem [{\citenamefont {Childress}\ and\ \citenamefont
  {McIntyre}(2010)}]{Childress10}%
  \BibitemOpen
  \bibfield  {author} {\bibinfo {author} {\bibfnamefont {L.}~\bibnamefont
  {Childress}}\ and\ \bibinfo {author} {\bibfnamefont {J.}~\bibnamefont
  {McIntyre}},\ }\bibfield  {title} {\enquote {\bibinfo {title} {Multifrequency
  spin resonance in diamond},}\ }\href@noop {} {\bibfield  {journal} {\bibinfo
  {journal} {Phys. Rev. A}\ }\textbf {\bibinfo {volume} {82}},\ \bibinfo
  {pages} {033839} (\bibinfo {year} {2010})}\BibitemShut {NoStop}%
\end{thebibliography}%


\begin{thebibliography}{12}%
\makeatletter
\providecommand \@ifxundefined [1]{%
 \@ifx{#1\undefined}
}%
\providecommand \@ifnum [1]{%
 \ifnum #1\expandafter \@firstoftwo
 \else \expandafter \@secondoftwo
 \fi
}%
\providecommand \@ifx [1]{%
 \ifx #1\expandafter \@firstoftwo
 \else \expandafter \@secondoftwo
 \fi
}%
\providecommand \natexlab [1]{#1}%
\providecommand \enquote  [1]{``#1''}%
\providecommand \bibnamefont  [1]{#1}%
\providecommand \bibfnamefont [1]{#1}%
\providecommand \citenamefont [1]{#1}%
\providecommand \href@noop [0]{\@secondoftwo}%
\providecommand \href [0]{\begingroup \@sanitize@url \@href}%
\providecommand \@href[1]{\@@startlink{#1}\@@href}%
\providecommand \@@href[1]{\endgroup#1\@@endlink}%
\providecommand \@sanitize@url [0]{\catcode `\\12\catcode `\$12\catcode
  `\&12\catcode `\#12\catcode `\^12\catcode `\_12\catcode `\%12\relax}%
\providecommand \@@startlink[1]{}%
\providecommand \@@endlink[0]{}%
\providecommand \url  [0]{\begingroup\@sanitize@url \@url }%
\providecommand \@url [1]{\endgroup\@href {#1}{\urlprefix }}%
\providecommand \urlprefix  [0]{URL }%
\providecommand \Eprint [0]{\href }%
\providecommand \doibase [0]{http://dx.doi.org/}%
\providecommand \selectlanguage [0]{\@gobble}%
\providecommand \bibinfo  [0]{\@secondoftwo}%
\providecommand \bibfield  [0]{\@secondoftwo}%
\providecommand \translation [1]{[#1]}%
\providecommand \BibitemOpen [0]{}%
\providecommand \bibitemStop [0]{}%
\providecommand \bibitemNoStop [0]{.\EOS\space}%
\providecommand \EOS [0]{\spacefactor3000\relax}%
\providecommand \BibitemShut  [1]{\csname bibitem#1\endcsname}%
\let\auto@bib@innerbib\@empty
\bibitem [{\citenamefont {Ionescu}\ and\ \citenamefont
  {Riel}(2011)}]{Ionescu11}%
  \BibitemOpen
  \bibfield  {author} {\bibinfo {author} {\bibfnamefont {A.~M.}\ \bibnamefont
  {Ionescu}}\ and\ \bibinfo {author} {\bibfnamefont {H.}~\bibnamefont {Riel}},\
  }\bibfield  {title} {\enquote {\bibinfo {title} {Tunnel field-effect
  transistors as energy-efficient electronic switches},}\ }\href@noop {}
  {\bibfield  {journal} {\bibinfo  {journal} {Nature}\ }\textbf {\bibinfo
  {volume} {479}},\ \bibinfo {pages} {329--337} (\bibinfo {year}
  {2011})}\BibitemShut {NoStop}%
\bibitem [{\citenamefont {Weber}\ \emph {et~al.}(1980)\citenamefont {Weber},
  \citenamefont {Schmid},\ and\ \citenamefont {Sauer}}]{weber1980localized}%
  \BibitemOpen
  \bibfield  {author} {\bibinfo {author} {\bibfnamefont {J.}~\bibnamefont
  {Weber}}, \bibinfo {author} {\bibfnamefont {W.}~\bibnamefont {Schmid}}, \
  and\ \bibinfo {author} {\bibfnamefont {R.}~\bibnamefont {Sauer}},\ }\bibfield
   {title} {\enquote {\bibinfo {title} {Localized exciton bound to an
  isoelectronic trap in silicon},}\ }\href@noop {} {\bibfield  {journal}
  {\bibinfo  {journal} {Phys. Rev. B}\ }\textbf {\bibinfo {volume} {21}},\
  \bibinfo {pages} {2401} (\bibinfo {year} {1980})}\BibitemShut {NoStop}%
\bibitem [{\citenamefont {Sauer}\ \emph {et~al.}(1984)\citenamefont {Sauer},
  \citenamefont {Weber},\ and\ \citenamefont {Zulehner}}]{sauer1984nitrogen}%
  \BibitemOpen
  \bibfield  {author} {\bibinfo {author} {\bibfnamefont {R.}~\bibnamefont
  {Sauer}}, \bibinfo {author} {\bibfnamefont {J.}~\bibnamefont {Weber}}, \ and\
  \bibinfo {author} {\bibfnamefont {W.}~\bibnamefont {Zulehner}},\ }\bibfield
  {title} {\enquote {\bibinfo {title} {Nitrogen in silicon: Towards the
  identification of the {1.1223-eV (A, B, C)} photoluminescence lines},}\
  }\href@noop {} {\bibfield  {journal} {\bibinfo  {journal} {Appl. Phys.
  Lett.}\ }\textbf {\bibinfo {volume} {44}},\ \bibinfo {pages} {440--442}
  (\bibinfo {year} {1984})}\BibitemShut {NoStop}%
\bibitem [{\citenamefont {Modavis}\ and\ \citenamefont
  {Hall}(1990)}]{modavis1990aluminum}%
  \BibitemOpen
  \bibfield  {author} {\bibinfo {author} {\bibfnamefont {R.~A.}\ \bibnamefont
  {Modavis}}\ and\ \bibinfo {author} {\bibfnamefont {D.~G.}\ \bibnamefont
  {Hall}},\ }\bibfield  {title} {\enquote {\bibinfo {title} {Aluminum-nitrogen
  isoelectronic trap in silicon},}\ }\href@noop {} {\bibfield  {journal}
  {\bibinfo  {journal} {J. Appl. Phys.}\ }\textbf {\bibinfo {volume} {67}},\
  \bibinfo {pages} {545--547} (\bibinfo {year} {1990})}\BibitemShut {NoStop}%
\bibitem [{\citenamefont {Iizuka}\ and\ \citenamefont
  {Nakayama}(2015)}]{iizuka2015first}%
  \BibitemOpen
  \bibfield  {author} {\bibinfo {author} {\bibfnamefont {S.}~\bibnamefont
  {Iizuka}}\ and\ \bibinfo {author} {\bibfnamefont {T.}~\bibnamefont
  {Nakayama}},\ }\bibfield  {title} {\enquote {\bibinfo {title}
  {First-principles calculation of electronic properties of isoelectronic
  impurity complexes in {Si}},}\ }\href@noop {} {\bibfield  {journal} {\bibinfo
   {journal} {Appl. Phys. Expr.}\ }\textbf {\bibinfo {volume} {8}},\ \bibinfo
  {pages} {081301} (\bibinfo {year} {2015})}\BibitemShut {NoStop}%
\bibitem [{\citenamefont {Ono}\ \emph {et~al.}(2019)\citenamefont {Ono},
  \citenamefont {Mori},\ and\ \citenamefont {Moriyama}}]{Ono18}%
  \BibitemOpen
  \bibfield  {author} {\bibinfo {author} {\bibfnamefont {K.}~\bibnamefont
  {Ono}}, \bibinfo {author} {\bibfnamefont {T.}~\bibnamefont {Mori}}, \ and\
  \bibinfo {author} {\bibfnamefont {S.}~\bibnamefont {Moriyama}},\ }\bibfield
  {title} {\enquote {\bibinfo {title} {High-temperature operation of a silicon
  qubit},}\ }\href@noop {} {\bibfield  {journal} {\bibinfo  {journal} {Sci.
  Rep.}\ }\textbf {\bibinfo {volume} {9}},\ \bibinfo {pages} {469} (\bibinfo
  {year} {2019})}\BibitemShut {NoStop}%
\bibitem [{\citenamefont {Pla}\ \emph {et~al.}(2012)\citenamefont {Pla},
  \citenamefont {Tan}, \citenamefont {Dehollain}, \citenamefont {Lim},
  \citenamefont {Morton}, \citenamefont {Jamieson}, \citenamefont {Dzurak},\
  and\ \citenamefont {Morello}}]{pla2012single}%
  \BibitemOpen
  \bibfield  {author} {\bibinfo {author} {\bibfnamefont {J.~J.}\ \bibnamefont
  {Pla}}, \bibinfo {author} {\bibfnamefont {K.~Y.}\ \bibnamefont {Tan}},
  \bibinfo {author} {\bibfnamefont {J.~P.}\ \bibnamefont {Dehollain}}, \bibinfo
  {author} {\bibfnamefont {W.~H.}\ \bibnamefont {Lim}}, \bibinfo {author}
  {\bibfnamefont {J.~J.~L.}\ \bibnamefont {Morton}}, \bibinfo {author}
  {\bibfnamefont {D.~N.}\ \bibnamefont {Jamieson}}, \bibinfo {author}
  {\bibfnamefont {A.~S.}\ \bibnamefont {Dzurak}}, \ and\ \bibinfo {author}
  {\bibfnamefont {A.}~\bibnamefont {Morello}},\ }\bibfield  {title} {\enquote
  {\bibinfo {title} {A single-atom electron spin qubit in silicon},}\
  }\href@noop {} {\bibfield  {journal} {\bibinfo  {journal} {Nature}\ }\textbf
  {\bibinfo {volume} {489}},\ \bibinfo {pages} {541--545} (\bibinfo {year}
  {2012})}\BibitemShut {NoStop}%
\bibitem [{\citenamefont {Rahman}\ \emph {et~al.}(2009)\citenamefont {Rahman},
  \citenamefont {Park}, \citenamefont {Boykin}, \citenamefont {Klimeck},
  \citenamefont {Rogge},\ and\ \citenamefont {Hollenberg}}]{rahman2009gate}%
  \BibitemOpen
  \bibfield  {author} {\bibinfo {author} {\bibfnamefont {R.}~\bibnamefont
  {Rahman}}, \bibinfo {author} {\bibfnamefont {S.~H.}\ \bibnamefont {Park}},
  \bibinfo {author} {\bibfnamefont {T.~B.}\ \bibnamefont {Boykin}}, \bibinfo
  {author} {\bibfnamefont {G.}~\bibnamefont {Klimeck}}, \bibinfo {author}
  {\bibfnamefont {S.}~\bibnamefont {Rogge}}, \ and\ \bibinfo {author}
  {\bibfnamefont {L.~C.~L.}\ \bibnamefont {Hollenberg}},\ }\bibfield  {title}
  {\enquote {\bibinfo {title} {Gate-induced g-factor control and dimensional
  transition for donors in multivalley semiconductors},}\ }\href@noop {}
  {\bibfield  {journal} {\bibinfo  {journal} {Phys. Rev. B}\ }\textbf {\bibinfo
  {volume} {80}},\ \bibinfo {pages} {155301} (\bibinfo {year}
  {2009})}\BibitemShut {NoStop}%
\bibitem [{\citenamefont {Silveri}\ \emph {et~al.}(2015)\citenamefont
  {Silveri}, \citenamefont {Kumar}, \citenamefont {Tuorila}, \citenamefont
  {Li}, \citenamefont {Veps\"al\"ainen}, \citenamefont {Thuneberg},\ and\
  \citenamefont {Paraoanu}}]{Silveri15}%
  \BibitemOpen
  \bibfield  {author} {\bibinfo {author} {\bibfnamefont {M.~P.}\ \bibnamefont
  {Silveri}}, \bibinfo {author} {\bibfnamefont {K.~S.}\ \bibnamefont {Kumar}},
  \bibinfo {author} {\bibfnamefont {J.}~\bibnamefont {Tuorila}}, \bibinfo
  {author} {\bibfnamefont {J.}~\bibnamefont {Li}}, \bibinfo {author}
  {\bibfnamefont {A.}~\bibnamefont {Veps\"al\"ainen}}, \bibinfo {author}
  {\bibfnamefont {E.~V.}\ \bibnamefont {Thuneberg}}, \ and\ \bibinfo {author}
  {\bibfnamefont {G.~S.}\ \bibnamefont {Paraoanu}},\ }\bibfield  {title}
  {\enquote {\bibinfo {title} {St\"uckelberg interference in a superconducting
  qubit under periodic latching modulation},}\ }\href@noop {} {\bibfield
  {journal} {\bibinfo  {journal} {New J. Phys.}\ }\textbf {\bibinfo {volume}
  {17}},\ \bibinfo {pages} {043058} (\bibinfo {year} {2015})}\BibitemShut
  {NoStop}%
\bibitem [{\citenamefont {Shevchenko}\ \emph {et~al.}(2005)\citenamefont
  {Shevchenko}, \citenamefont {Kiyko}, \citenamefont {Omelyanchouk},\ and\
  \citenamefont {Krech}}]{Shevchenko05}%
  \BibitemOpen
  \bibfield  {author} {\bibinfo {author} {\bibfnamefont {S.~N.}\ \bibnamefont
  {Shevchenko}}, \bibinfo {author} {\bibfnamefont {A.~S.}\ \bibnamefont
  {Kiyko}}, \bibinfo {author} {\bibfnamefont {A.~N.}\ \bibnamefont
  {Omelyanchouk}}, \ and\ \bibinfo {author} {\bibfnamefont {W.}~\bibnamefont
  {Krech}},\ }\bibfield  {title} {\enquote {\bibinfo {title} {Dynamic behavior
  of {Josephson-junction qubits: crossover between Rabi oscillations and
  Landau-Zener} transitions},}\ }\href@noop {} {\bibfield  {journal} {\bibinfo
  {journal} {Low Temp. Phys.}\ }\textbf {\bibinfo {volume} {31}},\ \bibinfo
  {pages} {569--576} (\bibinfo {year} {2005})}\BibitemShut {NoStop}%
\bibitem [{\citenamefont {Shevchenko}\ \emph {et~al.}(2010)\citenamefont
  {Shevchenko}, \citenamefont {Ashhab},\ and\ \citenamefont
  {Nori}}]{Shevchenko10}%
  \BibitemOpen
  \bibfield  {author} {\bibinfo {author} {\bibfnamefont {S.~N.}\ \bibnamefont
  {Shevchenko}}, \bibinfo {author} {\bibfnamefont {S.}~\bibnamefont {Ashhab}},
  \ and\ \bibinfo {author} {\bibfnamefont {F.}~\bibnamefont {Nori}},\
  }\bibfield  {title} {\enquote {\bibinfo {title} {{Landau-Zener-St\"uckelberg}
  interferometry},}\ }\href@noop {} {\bibfield  {journal} {\bibinfo  {journal}
  {Phys. Rep.}\ }\textbf {\bibinfo {volume} {492}},\ \bibinfo {pages} {1--30}
  (\bibinfo {year} {2010})}\BibitemShut {NoStop}%
\bibitem [{\citenamefont {Li}\ \emph {et~al.}(2013)\citenamefont {Li},
  \citenamefont {Silveri}, \citenamefont {Kumar}, \citenamefont {Pirkkalainen},
  \citenamefont {Veps\"al\"ainen}, \citenamefont {Chien}, \citenamefont
  {Tuorila}, \citenamefont {Sillanp\"a\"a}, \citenamefont {Hakonen},
  \citenamefont {Thuneberg},\ and\ \citenamefont {Paraoanu}}]{Li13}%
  \BibitemOpen
  \bibfield  {author} {\bibinfo {author} {\bibfnamefont {J.}~\bibnamefont
  {Li}}, \bibinfo {author} {\bibfnamefont {M.~P.}\ \bibnamefont {Silveri}},
  \bibinfo {author} {\bibfnamefont {K.~S.}\ \bibnamefont {Kumar}}, \bibinfo
  {author} {\bibfnamefont {J.-M.}\ \bibnamefont {Pirkkalainen}}, \bibinfo
  {author} {\bibfnamefont {A.}~\bibnamefont {Veps\"al\"ainen}}, \bibinfo
  {author} {\bibfnamefont {W.~C.}\ \bibnamefont {Chien}}, \bibinfo {author}
  {\bibfnamefont {J.}~\bibnamefont {Tuorila}}, \bibinfo {author} {\bibfnamefont
  {M.~A.}\ \bibnamefont {Sillanp\"a\"a}}, \bibinfo {author} {\bibfnamefont
  {P.~J.}\ \bibnamefont {Hakonen}}, \bibinfo {author} {\bibfnamefont {E.~V.}\
  \bibnamefont {Thuneberg}}, \ and\ \bibinfo {author} {\bibfnamefont {G.~S.}\
  \bibnamefont {Paraoanu}},\ }\bibfield  {title} {\enquote {\bibinfo {title}
  {Motional averaging in a superconducting qubit},}\ }\href@noop {} {\bibfield
  {journal} {\bibinfo  {journal} {Nat. Comm.}\ }\textbf {\bibinfo {volume}
  {4}},\ \bibinfo {pages} {1420} (\bibinfo {year} {2013})}\BibitemShut
  {NoStop}%
\end{thebibliography}%

\end{document}